 \def\dm15{$\Delta m_{15}$} \def\atrous{\`a trous}

\def\gs{\mathrel{\raise0.27ex\hbox{$>$}\kern-0.70em 
    \lower0.71ex\hbox{{$\scriptstyle \sim$}}}}
\def\ls{\mathrel{\raise0.27ex\hbox{$<$}\kern-0.70em 
    \lower0.71ex\hbox{{$\scriptstyle \sim$}}}}

\documentclass[12pt,preprint]{aastex}

\usepackage{epsfig} \usepackage{natbib} 

\slugcomment{Spectral Features of Type Ia Supernovae}

\shorttitle{Quantifying Spectral Features of Type Ia Supernovae}
\shortauthors{Wagers, Wang, Asztalos}

\begin{document}


\title{Quantifying Spectral Features of Type Ia Supernovae }


\author{A. Wagers$^1$, L. Wang$^1$ and S. Asztalos$^2$}
\affil{$^1$Department of Physics, Texas A\&M, College Station, Texas
  77843 \\$^2$ X-ray Instrumentation Associates, LLC, Hayward, CA
  94551 } \email{steve@xia.com}




\begin{abstract}
  We introduce a new technique to quantify highly structured spectra
  for which the definition of continua or spectral features in the
  observed flux spectra is difficult. The method employs wavelet
  transformation which allows the decomposition of the observed
  spectra into different scales. A procedure is formulated to define
  the strength of spectral features so that the measured spectral
  indices are independent of the flux levels and are insensitive to
  the definition of continuum and also to reddening. This technique is
  applied to Type Ia supernovae spectra, where correlations are
  revealed between the luminosity and spectral features. The current
  technique may allow for luminosity corrections based on spectral
  features in the use of Type Ia supernovae as cosmological probe.
\end{abstract}



\section{Introduction}
	
A major difficulty in analyzing spectroscopic data with highly blended
atomic lines is to quantify the strength of certain spectral features.
These spectral features are superimposed on a continuum so line
blending can make it difficult to reliably define the continuum level.
In the case of supernova spectra, the pre-nebular phase spectra
typically show P-Cygni profiles with both emission and absorption
components whereas the nebular phase spectra are dominated by broad
overlapping emission lines. The spectral features are therefore of
various widths and strengths, and neighboring features are heavily
blended. Further, the data typically contain observational noise, flux
calibrations errors and uncertainties in the amount of dust
extinction. The noise makes the definition of a continuum very
uncertain and accordingly the calculation of equivalent width becomes
unreliable. In many observations, in particular those at high
redshift, the observed supernova spectra are heavily contaminated by
host galaxy spectra. This affects severely the definition of line
depth.

For Type Ia supernovae, it is known that certain spectral line ratios
such as the Si II 5800/Si II 6150, and the Ca II H\&K lines are
sensitive to the intrinsic brightness of the supernova
\citet{1995ApJ...455L.147N}.  The measurement of the line strength is,
however, not trivial.  For instance, to measure the Si II 5800/Si II
6150, and the Ca II ratio, \citet{1995ApJ...455L.147N} employed a
simple approach by drawing straight lines at the local peaks of the
spectral features and measure the depth of the absorption minima from
the straight line. However, the location of the straight line and the
position of the line minimum are not easy to define in the presence of
observational errors. It is for this reason the ratio is derived only
for a number of very well observed local supernovae.

In this paper, the spectral features of Type Ia supernovae will be
analyzed instead through wavelet transformations. This technique
avoids many of the challenges mentioned above associated with
identifying line strengths. Here wavelet transformations are applied
to Type Ia supernovae spectra with the purpose to quantify the
spectral features for cosmological applications.

\begin{figure}
  \epsscale{1.0}

  \plotone{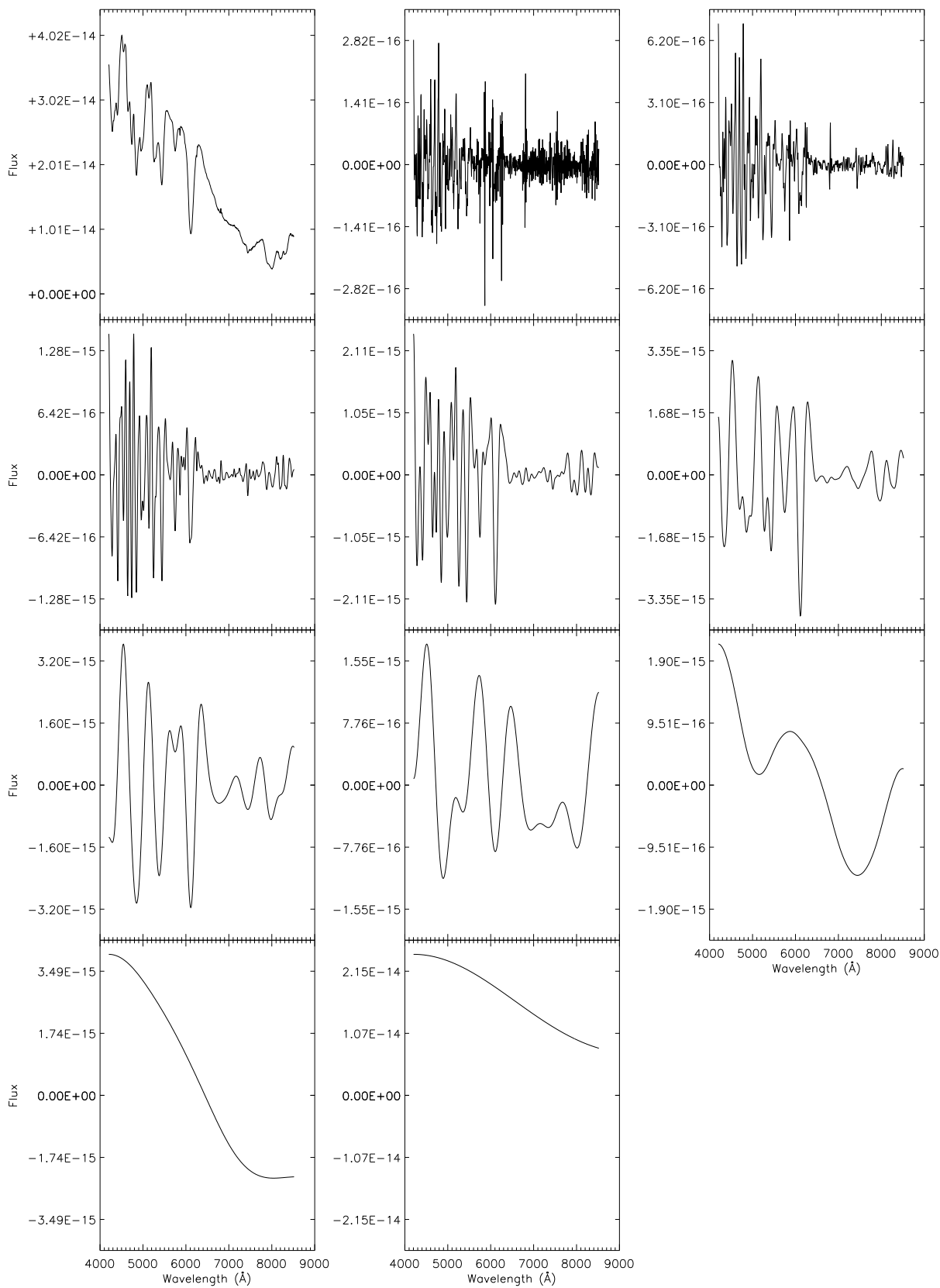}
  \caption{(a) The spectrum of SN 2001el, (b-j) the wavelet scales 1
    to 9, and (k) the smoothed array $c_p$ of the SN 2001el from top
    left to lower right. A sum of of scales (b) to (k) recovers the
    original spectrum (a). The mean fluxes of each of the wavelet
    scales (b) to (j) are identically zero.}\label{fig:fig1}
\end{figure}

\section{Spectral Features}

\subsection{Wavelet Transform Algorithm}

It has been previously demonstrated that the \atrous\ algorithm
\citet{1989SVHKMMT, 1995ASPC...77..279S, 1997ApJ...482.1011S} is a
very useful tool for studying spectral features. The transform is
carried out in direct space so artifacts related to periodicity do not
occur.  The reconstruction is trivial. The evolution of the transform
from one scale to the next is easy to follow and the interpretation of
the spectrum at each scale is straightforward.

The \`a trous wavelet uses a dyadic wavelet to merge non-dyadic data
in a simple and efficient procedure.  Assuming a scaling function
$\phi(x)$ (which corresponds to a low pass filter), the first
filtering is performed on the original data $\{c_0(k)\}$ by a twice
magnified scale leading to the $\{c_1(k)\}$ set. The signal difference
$\{c_0(k)\} - \{c_1(k)\}$ contains the information between these two
scales and is the discrete set associated with the wavelet transform
corresponding to $\phi(x)$. The operation is performed successively
and to obtain the wavelet scale {$w_j(k)$} at each scale $j$.  The
original spectrum $c_0$ can be expressed as the sum of all the wavelet
scales and the last smoothed array $c_p$:

$$c_0(k)\ = \ c_p(k)\ +\ \Sigma^p_{j=1} w_j(k)  $$

To demonstrate the basic features of the \`a trous wavelet
transformation, we show in Figure \ref{fig:fig1} the wavelet
transformation of a well observed supernova SN 2001el. The data were
obtained through the spectropolarimetry program at the Very Large
Telescope of the European Southern Observatory
\citet{2003ApJ...591.1110W}. The sampling step of the data is binned
to 5\AA.  The signal to noise ratio (SNR) of the data is everywhere
above 150 - this unusually high SNR is a result of the
spectropolarimetry observations.  The original data is show in Figure
\ref{fig:fig1}a, and the consecutive wavelet scales for j = 1, 9 are
shown in Figure \ref{fig:fig1}b to \ref{fig:fig1}j. Figure
\ref{fig:fig1}k represents the smoothed array $c_p$: given the
spectral range of SN 2001el the \`a trous wavelet cannot generate more
than 10 wavelet scales. Further note that each of the individual
wavelet scales have zero mean.  It can be seen that at small scales
the wavelet is dominated by observational noise and the supernova
signal starts to become significant only for $j \ge 3$, and the broad
spectral wiggles associated with the supernova dominate the wavelet
scales of j = 5, 6, and 7. The supernova spectral features are
typically a few hundred \AA\ wide and are effectively isolated in the
decomposed spectra.

\begin{figure}
  \epsscale{1.0}

  \plotone{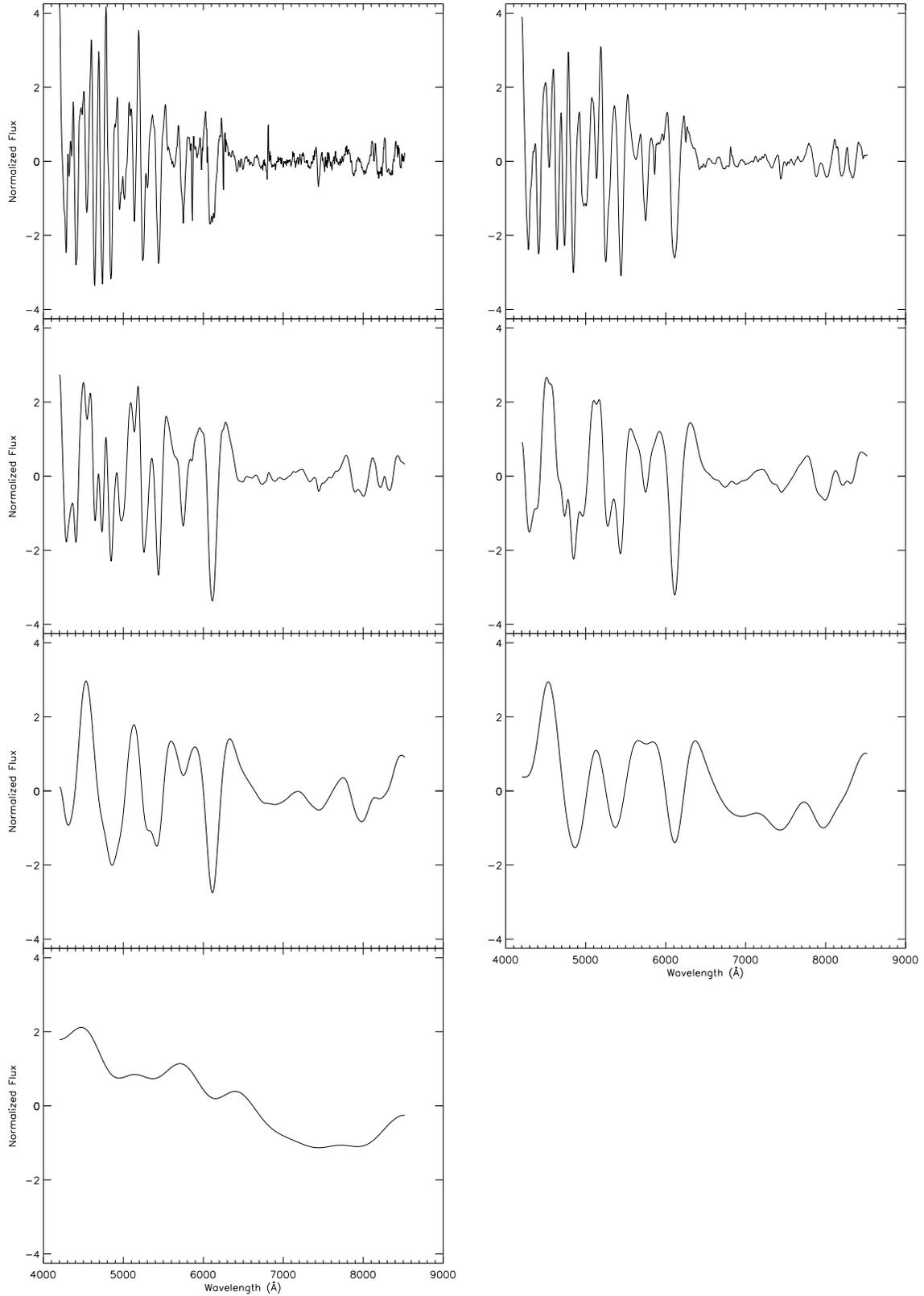}
  \caption{Wavelet sums for SN 2001el over (a) scales 1 to 3, (b) 2 to
    4, (c) 3 to 5, (d) 4 to 6, (e) 5 to 7, (f) 6 to 8 and (g) 7 to 9.
    Running combinations of the different scales captures features of
    varying width.}\label{fig:fig2}
\end{figure}

The spectral features of a supernova can be better described by a
blend of several wavelet scales. For this reason, we can calculate the
sum of more than one scales to reflect the existence of features of
various width: $$W_{\{l\}} \ = \ \Sigma_{j\in{\{l\}}} w_j,$$ where
$\{l\}$ is a subset of wavelet scales. Examples of these spectra are
shown in Figure \ref{fig:fig2} for SN 2001el.

\subsection{Normalization of Spectral Features}

The wavelet scales, having the units of the original flux spectrum,
need to be normalized to construct quantities that measure the
strength of the spectral features that do not depend on the absolute
flux level of the spectrum. There undoubtedly is more than one way to
normalize the scales. The simplest approach is to normalize all the
wavelet scales by dividing them by the smoothed array $c_p$.  This
approach is simple and will certainly work fine for data without host
galaxy contamination. For data with host galaxy contamination, or
those with poor background subtraction, this approach introduces
systematic errors to the normalized scales.

In our approach, the normalized wavelet scale is defined using the
standard deviations of the spectral features from any given wavelet
scale:
\begin{eqnarray} {\hat W_{\{l\}}} (\lambda)\ &=& \ W_{\{l\}}(\lambda)/
  \sqrt{\Sigma_{\lambda_1}^{\lambda_2} W_{\{l\}}^2(\lambda)/N_{12}}\\
  &=& W_{\{l\}}(\lambda)/ \sigma_{\{l\}},\label{eq:sigma}
\end{eqnarray}
where $N_{12}$ is the number of data points between $\lambda_1$ and
$\lambda_2$.  The mean and standard deviation of ${\hat W_{\{l\}}}$
are zero and 1, respectively.  This is effectively a
self-normalization that exploits only the intrinsic properties of the
wavelet scales involved. Host galaxy contamination (which does
strongly affect $c_p$) would not have a significant effect in this
context.

The spectral index $X_j$ of any feature between $\lambda_a$ and
$\lambda_b$ at a given scale $j$ is defined by averaging the
normalized wavelet scale {$\hat W_j$}:
$$X_{\{l\}}\ = \ \Sigma_{\lambda_a}^{\lambda_b}\hat W_{\{l\}}(\lambda)/N_{ab}, $$
with $N_{ab}$ being the bin size in the wavelength region $\lambda_a$
and $\lambda_b$. $X_j$ defines a normalized number which measures the
strength of the spectral features in the normalized wavelet scale
$\hat W_j$ between $\lambda_a$ and $\lambda_b$. Alternatively, one can
also calculate the power $P_j$ between $\lambda_a$ and $\lambda_b$ for
wavelet scale $j$:
$$P_{\{l\}}\ = \ \Sigma_{\lambda_a}^{\lambda_b} \hat W_{\{l\}}^2(\lambda)/N_{ab}. $$
$P_{\{l\}}$ and $X_{\{l\}}$ contain the same information.  In this
study, we will focus on $X_{\{l\}}$.

One obvious advantage to using wavelet scales to estimate spectral
feature strengths is that they do not depend on the definition of the
spectral continua.  Furthermore, since they can be estimated locally
around a spectral feature, spectral indices are useful in minimizing
uncertainties due to errors in spectral flux calibrations.  Similarly,
the spectral indices as defined here are less sensitive to errors of
background subtraction, which is usually one of the dominate sources
of uncertainty, especially in the studies of high redshift supernovae.

\subsection{Normalization Spectral Features of SN Ia}

The wavelet technique is particularly well-suited for studying
scattering-dominated spectra of expanding atmospheres with P-Cygni
spectral features: the net flux of the P-Cygni feature is usually
close to zero. Wavelet decomposition is consistent with this as the
mean flux is zero for the various wavelet scales. Wavelet transforms
thus makes it easy to separate emission and absorption components of a
spectrum in a mathematically robust way.

\begin{figure}
  \epsscale{1.0}
  \plotone{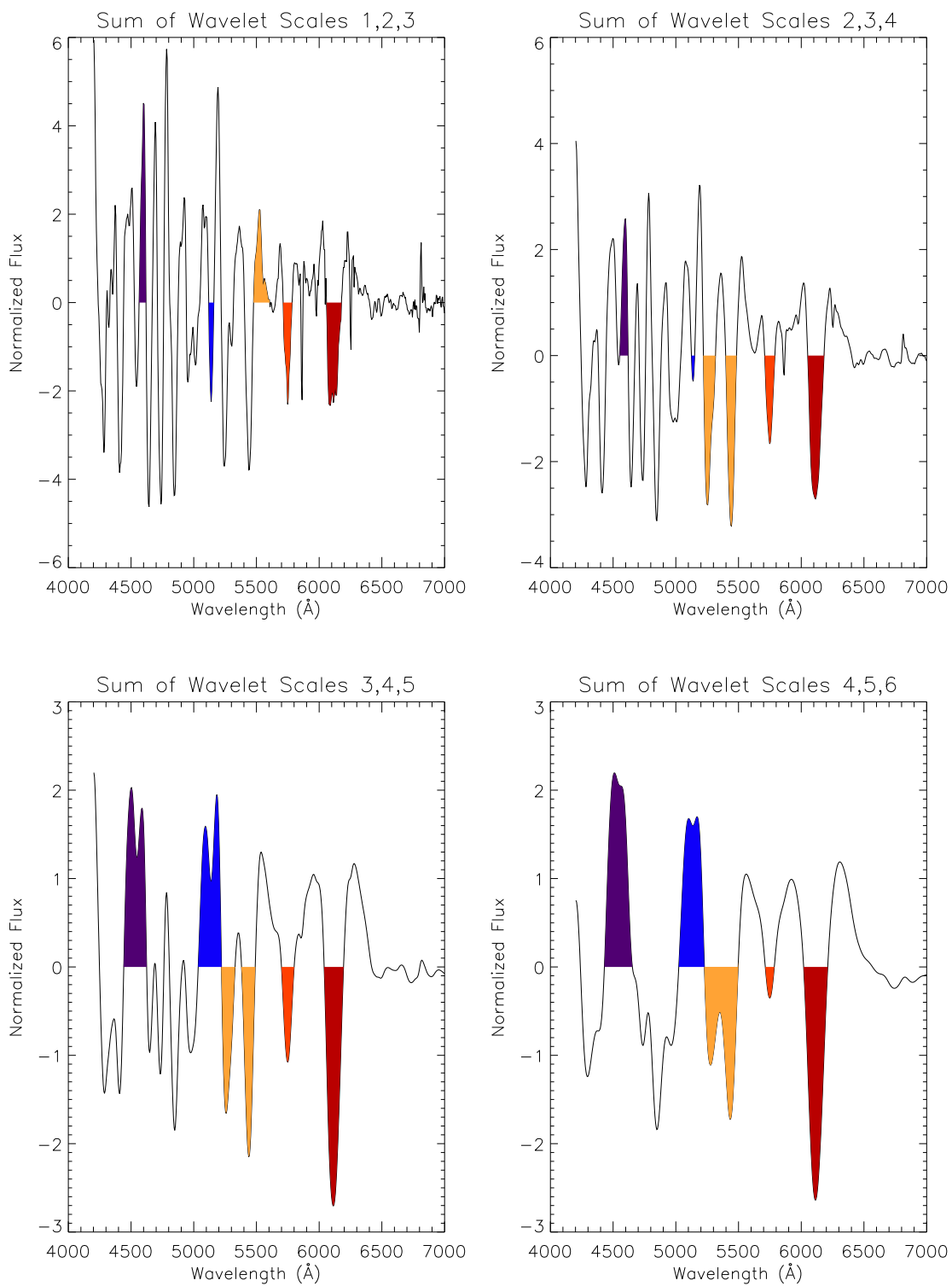}
  \caption{Spectral features for Type Ia supernovae. (a): {l} =
    {1,2,3}.  (b): {l} = {2,3,4}, (c) {l} = {3,4,5}, and (d) {l} =
    {4,5,6}.  The spectral features are well resolved in (c) and
    (d). }\label{fig:fig3}
\end{figure}

In this study, the supernova spectra are first decomposed into various
scales as described in the above section. In addition, to reflect the
fact the spectral features are a blend of different scales, the sum of
the wavelet scales 3, 4, and 5 are used as the primary spectrum for
the analysis of spectral features (though other scales have also been
analyzed). All the decomposed spectra are normalized in a similar way
as given in Equation \ref{eq:sigma}. To derive quantities that are
less sensitive to errors of flux calibration, we need to restrict
calculation of the normalization factor to a small wavelength region
and yet to have large enough spectral coverage so that the feature
strengths will not be affected by boundary. In this study, the spectra
are divided into four regions: (A) 5500 to 6500 \AA, (B) 4985 to 5985
\AA, (C) 4850 to 5450 \AA, and (D) 4250 to 5200 \AA; the variance in
each of these sections of spectra is calculated and used as the
normalization factor. Interesting features include the Silicon II
lines at 635.5 and 580.0 nm in region (A), the Silicon II line around
548.5nm in region (B), and the two strong peaks at 510.0 nm and 450.0
nm, in region (C) and (D), respectively.  These five spectral features
are shown in Figure \ref{fig:fig3} and are the main focus of
subsequent analyses.

\section{Biases and Errors}\label{sec:ba}

In practice, the observed data contain noise and estimates of $X_j$
can be biased. The noise affects $X_j$ in two ways: First, when the
noise is large, its effect can propagate to all the wavelet scales and
become a significant component at the wavelet scale of interest.
Secondly, it changes the normalization factor when calculating $ \hat
W_j$ - data with larger noise can be systematically biased to give a
larger normalization factor because the additional power from shot
noise. This bias is usually not a problem for high SNR
ratio data, but can be significant for data with a low SNR
ratio. The correction factor $\Pi(j)$ for scale j is defined as:

\begin{equation}
\sigma_0(j) \ = \sigma(j)\Pi_{1j},
\end{equation}\label{eq:chi}

where $\sigma_0(j)$ is the variance at the jth scale in the ideal case
of no photon shot noise.

Typically, published spectroscopic data do not have associated noise
spectra. One instead has to rely on the flux spectrum to estimate the
noise levels.  A major advantage of wavelet transformation is that it
allows estimates of the noise characteristics based on the spectral
data itself. If we assume that all the continuum or spectral features
are much broader than the data sampling step, the spectral
fluctuations of the wavelet scale with $j=1$ should then represent
mostly shot noise.  This is generally reasonable as can be seen in
Figure \ref{fig:fig1} (b) for the spectrum of Type Ia supernova
2001el.

Recognizing that smaller wavelet scales contain more information of
the noise property than larger wavelet scales, we can define the
spectral quality index (SQI) of the normalized wavelet scale $\{l\}$
as the variance ratio of normalized scale $\{l\}$ and the lowest
normalized scale $\{1\}$:

\begin{equation}
  \rho_{{\{l\}}\{1\}}\ = \ \Sigma_{\lambda1}^{\lambda2}
 \hat  W_{\{l\}}^2/\Sigma_{\lambda1}^{\lambda2}\hat W_{\{1\}}^2,
\end{equation}\label{eq:rho}

where the braces $\{\}$ reflect that the various quantities are
actually sums of various wavelet scales. Specifically, $\hat
W_{\{1\}}$ is the normalized sum of three wavelet scales, where
$\{l\}$ \ = \ 1,2,3 in this instance. The SQI measures the relative
importance of noise levels in estimating of the spectral feature
index. It can be calculated directly from the decomposed spectra
without an error spectrum. Note that SQI is a quantity that can be
localized to certain wavelength intervals.

For a given spectrum \{$c_i$\}, the dependence of the wavelet spectral
indices $X_{\{l\}}$ and the correction factor $\Pi$ on SQI can be
estimated through Monte-Carlo simulations.

\subsection{Dependence of Spectral Features on Observational Noise}

Monte Carlo simulations are required to quantify the dependence of $X$
indices on observational noise.  The characteristic parameter of the
noise is the SQI defined in Equation \ref{eq:rho} - the ratio of
spectral variance of combined wavelet scales ${l}\ = \ {3,4,5}$ to
that of the combined scales ${l}\ = \ {1,2,3}$. To perform such these
simulations one needs a series of noise free spectra of supernova
spectra. The spectropolarimetry program at the ESO VLT has acquired
several high quality spectra of SN Ia with SNR ratios around 150
(2003ApJ...591.1110W). Spectra of SN~2001V and SN~2001el from the
spectropolarimetry program will be used in this simulation to quantify
the relations between $X$ and SQI.

\begin{figure}
  \epsscale{0.8}

  \plotone{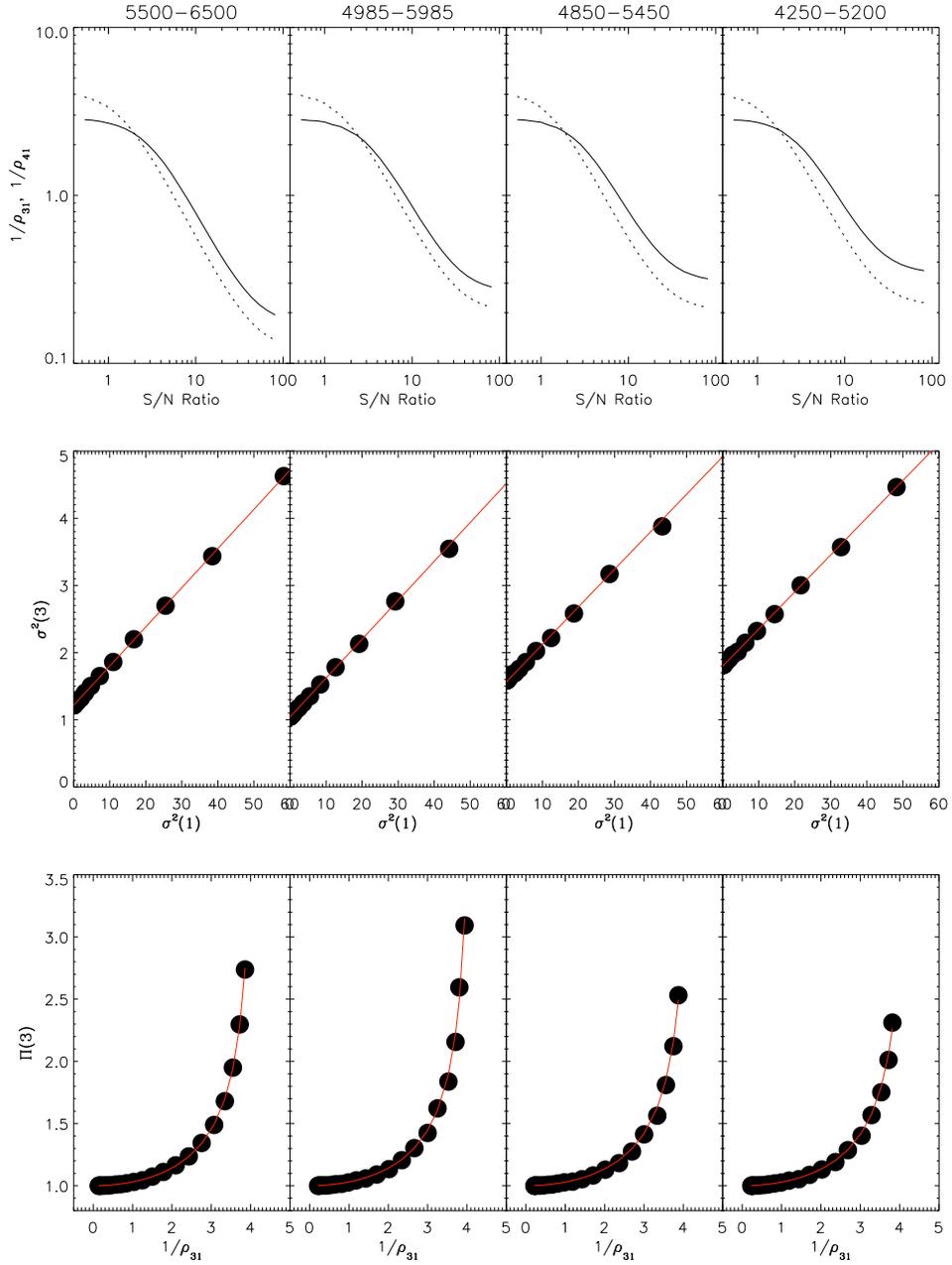}
  \caption{(Top) The relation between the SQI and the input SNR
    ratios. The solid lines show SQI of fourth wavelet scale and the
    dashed line the third. (Middle) The relation between the variance
    of the third and first wavelet scales. The effect of a large
    $\sigma(1)$ propagates linearly to larger wavelet scales. (Bottom)
    The correlation of the bias correction factor and SQI for the
    third wavelet scale.  The SNR was varied in all cases via the
    addition of noise in the Monte Carlo
    simulations. }\label{fig:fig4}
\end{figure}

In the example shown in Figure \ref{fig:fig4}, various levels of
Poisson noises were added to the spectrum of SN 2001el at day +1. The
noise is added to the spectra, which are then transformed to various
wavelet scales and the various $X$ indices are calculated. The top
panel in Figure \ref{fig:fig4} shows the relation between $\rho$ and
the assumed SNR ratio with the addition of Poisson noise. The SQI is
calculated in the wavelength intervals of 550.0 nm to 650.0 nm,
498.5nm to 598.5 nm, and 425.0 to 520.0 nm.  It can be seen that
$\rho$ is correlates well with the SNR ratio of the input data:
reducing the SNR ratio decreases $\rho$.  This confirms that the SQI
can effectively capture effect the photon shot noise, and can be used
to quantify the noise level of the data.

The variances used to normalized the spectra at the various wavelet
scales are clearly correlated. This is shown in the middle panel of
Figure \ref{fig:fig4}, where the data exhibit nearly identical slopes
over the different wavelength regions.  A linear relationship between
$\sigma^{2}(1)$ and $\sigma^{2}(3)$, and between $\sigma^{2}(1)$ and
$\sigma^{2}(4)$ is assumed for the fits. The slopes $\gamma_{1j}$
extracted from these fits are given in Table \ref{tab:tab1}.

The bottom panel in Figure \ref{fig:fig4} clearly demonstrates how the
correction factor $\Pi$ increases dramatically for $\rho$ approaching
2.82 (which corresponds to a SNR ratio of below 1 per 0.5
nm bin). This implies that the spectral features are dominated by the
noise, hence it becomes impossible to extract the spectral indices
reliably.

The correction factor for $\rho$ can be fit well with a function
\begin{equation}
  \Pi_{1j}\  =\  \sqrt{(1-\gamma_{1j} \rho_{1j}^2)},
\end{equation}\label{eq:pi}
with the relevant coefficients taken from Table \ref{tab:tab2} for the
various lines.

\subsection{Bias corrections}

\begin{figure}
  \epsscale{1.0}

  \plotone{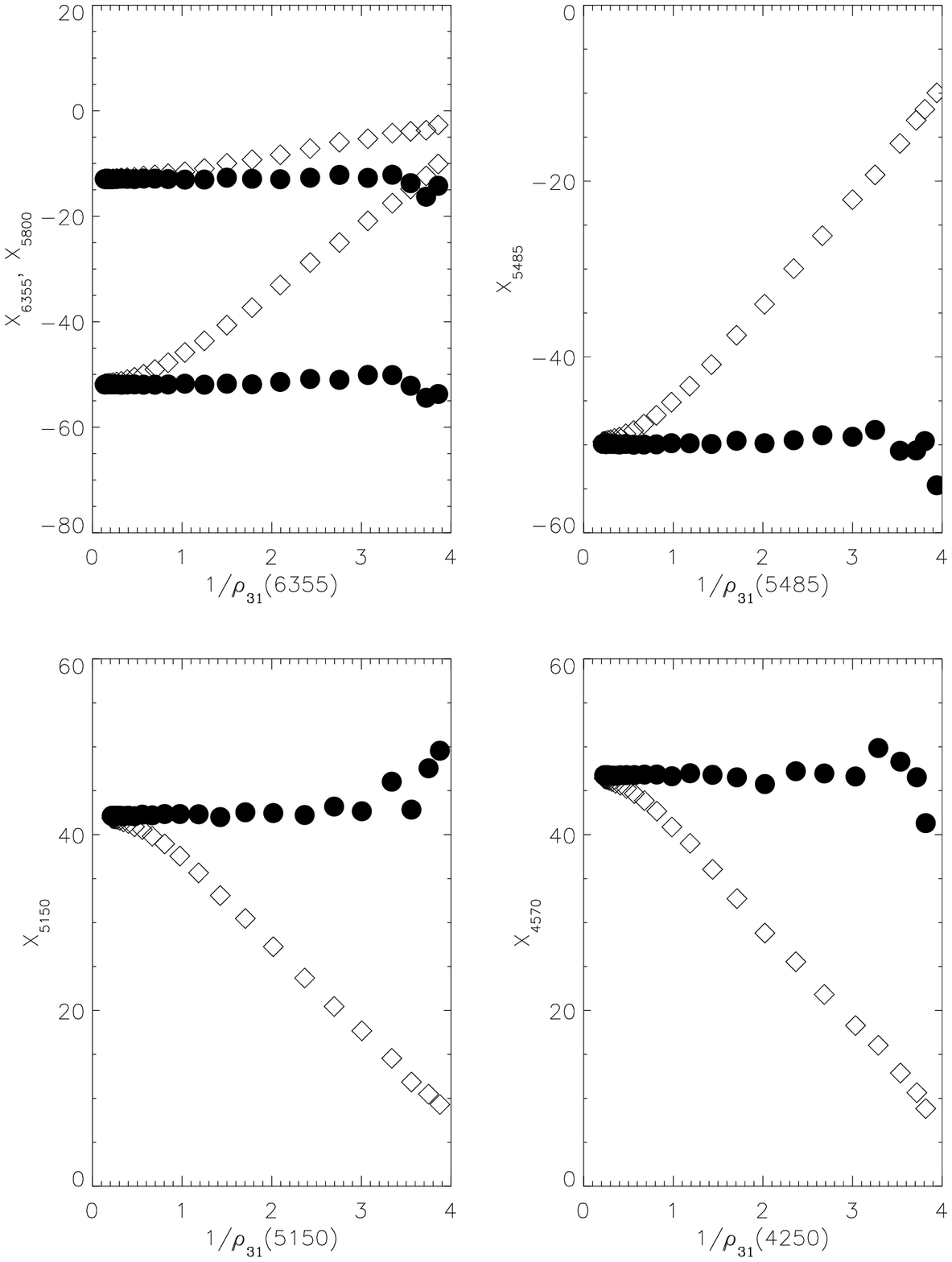}
  \caption{Line indices corrected for $\rho$ dependence for important
    SN~Type~1a spectral features. The $X$ indices are derived from the
    sum of of wavelets {3,4,5} of SN~2001el 1 day past optical
    maximum.  }\label{fig:fig5}
\end{figure}

The various $X$ indices for the spectral features are derived from the
Monte-Carlo simulation of data with different SQI. As shown in Figure
\ref{fig:fig5}, the $X$ indices (shown as open squares apparently
suffer strong bias when the data are noisy.  The various $X$ indices
after $\Pi$ corrections are shown in Figure \ref{fig:fig5}. The effect
is generally small for high SNR ratio data, but becomes
important for data with low SNR ratio. In any case the
bias is effectively removed by applying the correction factor $\Pi$.

\subsection{Error Estimates of the Spectral Indices}

\begin{figure}
  \epsscale{1.0}

  \plotone{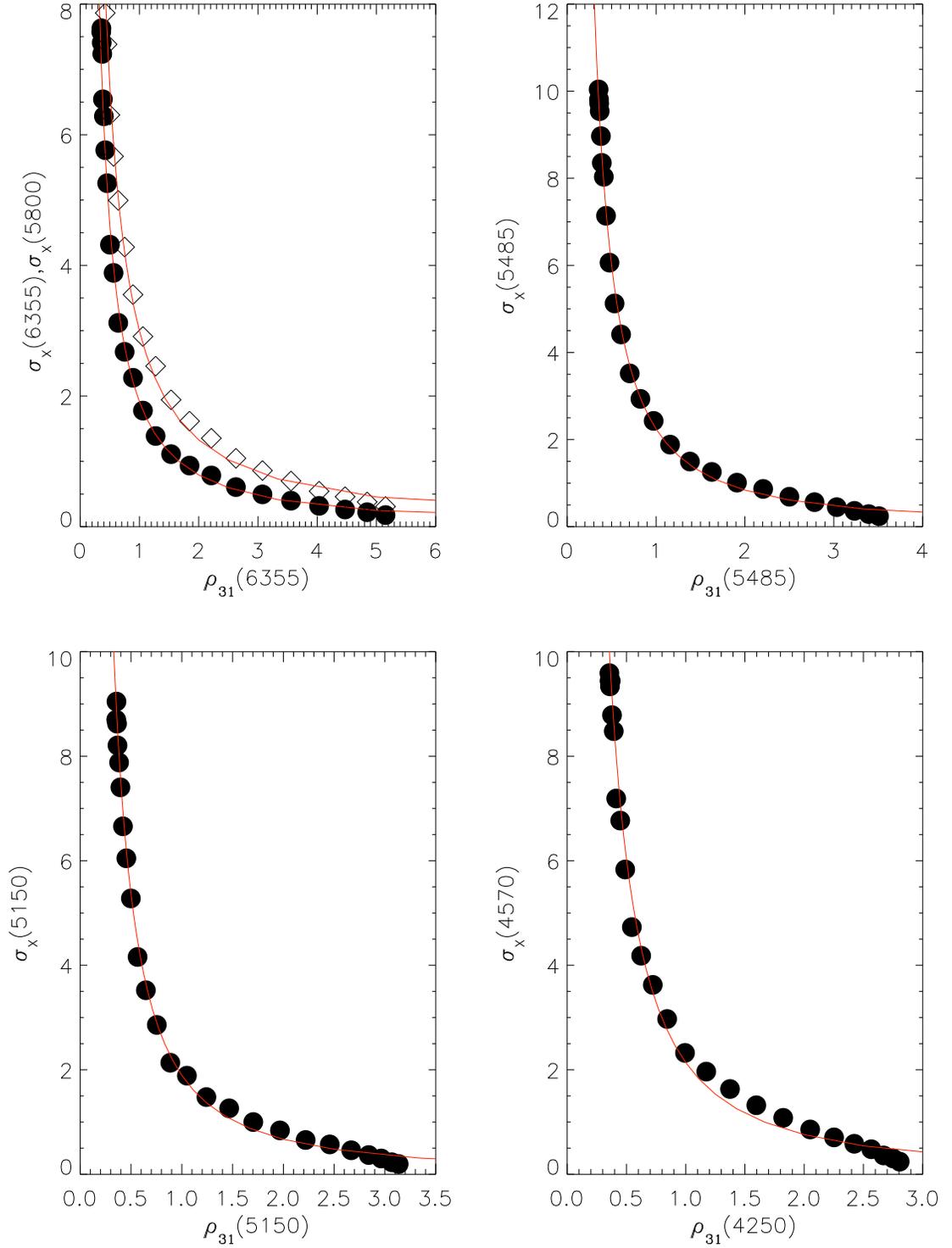}
  \caption{Errors of the $X$ indices as functions of $\rho$ for
    SN~2001el at optical maximum.  }\label{fig:fig6}
\end{figure}

Assuming photon shot noise, the Monte Carlo simulations also give
error estimates for the $X$ indices. The errors as a function of
$\rho$ are shown in Figure \ref{fig:fig6}. These errors are fitted
with a function of the form:
\begin{equation}
  \sigma_{X}\ = \ {\eta \rho^\psi}, 
\end{equation}\label{eq:sigchi}

and the relevant coefficients $\eta$ and $\psi$ are shown in Table
\ref{tab:tab2}.  Simulations were performed for all of the SN~2001V
and SN~2001el spectra and it was found that in all cases the bias can
be well corrected.  Note that due to the lack of a completely
noise-free SN~Ia spectrum, at extremely high SNR ratio (such as those
that are higher than or comparable with the signal to noise ratio of
the SN 2001el spectra as used in the simulation) the Monte Carlo
simulations do not give correct estimates of the errors. Such cases
are unlikely to be relevant as in such situations, the errors are
likely to be dominated by calibration systematics rather than shot
noise. The error function given above will be used for all cases
here. As can be seen in Figure \ref{fig:fig6}, the above expression
gives an excellent description of the dependence when the errors are
described by $\rho$.

The $\rho$ dependence of the $X$ indices and their errors have a weak
dependence among the different varieties of supernovae and the epoch
of the supernovae.

\begin{table}
  \caption{The Coefficients for the Dependence of $X$ on Data Errors}\label{tab:tab1}
  \begin{tabular}{lr||llll|l||llll|l}
    \hline

    SN & Day & $\gamma_{31}(A)$ & $\gamma_{31}(B)$ & $\gamma_{31}(C)$ &$\gamma_{31}(D)$& mean& $\gamma_{41}(A)$ & $\gamma_{41}(B)$ & $\gamma_{41}(C)$ & $\gamma_{41}(D)$ & mean \\
    \hline
    \hline
    01V   & $-8$  & 0.1132     & 0.1116           & 0.1183        &0.1144   & 0.1144     &       0.0263           & 0.0240 & 0.0273 & 0.0259  & 0.0258 \\
    01el 	& $-4$	& 0.1220     & 0.1236		& 0.1229	&0.1224   & 0.1227     &       0.0567	         & 0.0587 & 0.0566 & 0.0573  & 0.0573 \\
    01el  & +1    & 0.1220     & 0.1236           & 0.1229        &0.1224   & 0.1227            & 0.0581           & 0.0583 & 0.0552 & 0.0551 & 0.0567\\
    01el  & +9   & 0.1251	     & 0.1273		& 0.1250	&0.1240	  & 0.1254	      & 0.0585           & 0.0607 & 0.0599 & 0.0577 & 0.0592 \\
    \hline
  \end{tabular}
\end{table}

\begin{table}
  \caption{The Coefficients for the Errors of $X$}\label{tab:tab2}
\tiny{
    \begin{tabular}{lr||ll|ll||ll|ll|ll||}
      \hline
      SN & Date & $\eta(6150) $ & $\psi(6150)$ & $\eta(5800)$ & $\psi(5800)$ & $\eta(5485)$ & $\psi(5485)$ & 
      $\eta(5150)$ & $\psi(5150)$  & $\eta(4250)$ & $\psi(4250)$  \\ 
      \hline
      \hline
      2001V   & $-8$  &-0.02486    & 0.0469           &-0.0156        &0.0293   & -0.0203           & 0.0601   &
      -0.0373 & 0.0545 & -0.0179   & 0.0453 \\
      2001el 	& $-4$	&-0.00625     & 0. 0367          &-0.00628        &0.05429   & -0.02449            & 0.05892   &
      -0.01626 & 0.06382 & -0.01165   & 0.05248 \\
      2001el  & +1    &0.133    & 1.248           &0.212        &1.188   & 0.160           & 1.412   & 
      0.176 & 1.533 & 0.157   & 1.531  \\
      2001el  & +9   & 0.159    & 1.276           &0.194        &1.282   & 0.234           & 1.279   &
      0.203 & 1.442 & 0.177   & 1.672 \\
      \hline
      \hline
      2001el & +1     &0.145    & 1.194           &0.265        &1.322   & 0.175           &1.295    &
      0.181  & 1.500  & 0.216   & 1.408 \\

      \hline
    \end{tabular}
  }
\end{table}

\textbf{Procedure for Bias and Error Estimation\\} The procedure for
removing bias and estimating errors from noisy supernovae spectra is
premised on extracting correction factors from a supernova with a
large SNR. Here we enumerate a correction recipe using wavelet scales
$l = {3,4,5}$ with SN~2001~el as our reference spectrum.

\begin{enumerate}
\item Compute the ratio of the sum of the squares of wavelet scales
  3,4,5 and 1,2,3 ($\rho_{31}$ in Eq. \ref{eq:rho}) for SN~2001~el
$$\rho_{31} = \frac{ \hat W^2_{\{3,4,5\}}}{\hat W^2_{\{1,2,3\}}}. $$
\item Degrade the SNR ratio of SN~2001~el in multiple
  steps with the addition of Poisson noise. Compute
  $\sigma_{\{1,2,3\}}^2$ and $\sigma_{\{3,4,5\}}^2$ at each step.
\item Extract $\gamma_{31}$ by assuming a linear relationship between
  $\sigma_{\{1,2,3\}}^2$ and $\sigma_{\{3,4,5\}}^2$ (see middle panel
  of Figure \ref{fig:fig4}). 
$$\sigma^2_{\{3,4,5\}} = \beta + \gamma_{31} * \sigma^2_{\{1,2,3\}}, $$

where $\sigma_{\{l\}}$ is defined in Eq. \ref{eq:sigma}.
\item Repeat each of the above steps over all regions of interest to
  extract a mean value for $\gamma_{31}$.  Equivalently, use the
  values for $\gamma_{31}$ by consulting Table \ref{tab:tab1}.
\item Using equation \ref{eq:pi}, compute $\rho_{31}$ values for each
  SNe having typical values of the SNR ratio. 
\item Compute the correction factor $\Pi_{31}$ using this value of
  $\rho_{31}$ and the value of $\gamma_{31}$ computed for SN~2001~el
  $$\Pi_{31} = \sqrt{1-\gamma_{31} \rho^2_{31}}.$$
\item The spectral index of any supernovae feature can be corrected
  for bias by dividing the uncorrected value by $\Pi_{31}$ as in
  Eq. \ref{eq:chi}.
$$X_{corr} = \frac {X}{\Pi_{31}}.$$
\item The error bars are determined from the same set of
  simulations. With $\sigma_{X}$ defined as in Equation
  \ref{eq:sigchi} construct the relationship
$$log_{10}(\sigma_{X}) = \eta + \psi log_{10}(\rho_{31}),$$
where all quantities refer to a SNe with a large SNR ratio
(e.g. SN~2001~el). Equivalently, use the values for $\gamma_{31}$ by
consulting Table \ref{tab:tab2}.
\item With these values $\eta$ and $\psi$ compute the variance in
  the spectral index for a SN with a typical value of SNR as
  $$\sigma_{X} = 10^{\eta} \rho_{31}^{\psi}.$$
\end{enumerate}

\section{Applications to Type Ia Supernovae}\label{sec:appl}

Spectral indices lend themselves to a quantitative analysis of the
temporal and magnitude evolution of the spectral lines.
\citet{1995ApJ...455L.147N} measured the {\it ratio} of the depths of
the \ion{Si}{2} 6150 \AA\ and 5750 \AA\ features and established
correlations with $\Delta m_{15}$. Other studies of these and other
spectral features have adopted slightly more elaborate procedures
based on equivalent \citet{2006MNRAS.370..299H} and pseudo equivalent
widths (EW)
\citet{2004623,2007A&A...470..411G,2009ApJ...695..135A,2009PASP..121..238B}
to study these and other absorption features. Recently,
\citet{2008A&A...492..535A,2007A&A...469..645S} have used wavelets
coupled with the pseudo-EW technique to study a \ion{Si}{2} absorption
feature.  Emission features have received less attention than
absorption features and usually involve a distinct procedure from the
absorption features \citet{1995ApJ...455L.147N,2006ApJ...647..513B}.
Recently, \citet{2009arXiv0905.0340B} described a variance of the
above methods wherein absorption and emission features in a training
set of spectra are studied to extract the optimal flux ratio to
$\Delta m_{15}$ correlation. This ratio is then applied to correct the
magnitudes of other supernovae within a validation set.

The wavelet technique developed here differs in several important
respects from those described in the preceding paragraph. Our
methodology is premised on the existence of one (or more) very high
SNR spectra. Spectral line strengths are first extracted for this high
SNR spectrum from a combination of intermediate wavelet
scales. Excluding the lowest and highest reduces the effects of noise
and the continuum, respectively. Working with wavelets from a high SNR
spectrum allows corrections to be made to lower SNR spectra. Perhaps
the most salient difference is that performing our analysis entirely
in wavelet space permits us to avoid definition of the continuum (the
mean of the wavelet scales is zero and integration is performed from
one zero on the leading to the next zero on the trailing edge of a
feature). Consequently, we are able to work directly with line
strengths of the features themselves, not their ratios. Lastly, this
technique permits absorption and emission lines to be treated
democratically. It is our expectation that the wavelet method gives to
a more robust measure of line strength. In Sections \ref{epoch} and
\ref{delta} we apply this technique to data described in the
subsequent section.

\subsection{Data Sample}

The supernovae included in this study are given in Table
\ref{tab:tab3}.  A large number of low-z spectra (z $<$ 0.1) were
collected from libraries that are publicly available, such as the
SUSPECT Supernova
Database\footnote{http://bruford.nhn.ou.edu/~suspect/index1.html} and
the Center for Astrophysics Supernova
Archive\footnote{http://www.cfa.harvard.edu/supernova/SNarchive.html},
as well as other SNe that are available in the literature.  The
spectra are corrected by the host galaxy redshift but no dust
extinction correction is applied. The original wavelength coverages,
step sizes and SNR ratios of these spectra are vastly different. In
our analysis, all the spectra are first rebinned to 5 \AA\ sampling
step for convenience.  After wavelet decomposition was performed the
spectra were checked for edge effects that would distort calculations,
the affected spectra were removed.

\begin{deluxetable}{lcccccccc}
\tabletypesize{\tiny}
\tablewidth{500pt}
\tablehead{\colhead{SNe} & \colhead{$\Delta m_{15}$\tablenotemark{a}} & \colhead{Branch Subtype\tablenotemark{b}} & \colhead{$X_{6150}$} & \colhead{$X_{5750}$} & \colhead{$X_{5485}$} & \colhead{$X_{5150}$} & \colhead{$X_{4570}$} & \colhead{Spectra Source}}
\tablecaption{The Spectroscopic Sample of SNe Ia}\label{tab:tab3}
\startdata



1981B & 1.125(0.010) & BL & -56.645(0.157) & -10.339(0.298) & -48.941(0.245) & 42.956(0.175) & 43.306(0.184) & 1 \\
1983G & 1.37(0.01)\tablenotemark{b} & \nodata  & -59.818(0.649) & -13.890(3.411) & \nodata\tablenotemark{k} & 49.532(0.483) & \nodata\tablenotemark{k} & 2 \\
1984A & 1.294(0.063) & BL & -63.615(0.974) & -5.166(1.167) & -39.319(0.489) & 45.827(0.573) & 43.895(0.781) & 3 \\
1986G & 1.643(0.022) & CL & -49.109(0.295) & -20.165(0.534) & -32.227(0.456) & 40.231(0.386) & 51.494(0.521) & 4 \\
1989B & 1.262(0.017) & CL & -54.202(0.486) & -13.218(0.428) & -47.078(0.424) & 41.870(0.216) & 46.161(0.318) & 5,6 \\
1990N & 1.138(0.024) & CN & -53.751(0.112) & -9.683(1.763) & \nodata\tablenotemark{k} & 39.630(0.312) & \nodata\tablenotemark{k} & 7,8 \\
1991T & 0.986(0.009) & SS & -52.505(0.274) & -2.976(0.499) & -29.806(0.829) & 34.860(0.159) & 21.783(0.474) & 9,10,11 \\
1991bg & 1.857(0.125) & CL & -46.461(0.324) & -25.588(0.582) & -22.972(0.382) & 24.081(0.436) & 51.622(0.367) & 12,13,14 \\
1992A & 1.320(0.015) & BL & -54.040(0.242) & -16.239(0.517) & -46.671(0.447) & 41.067(0.300) & 36.031(0.336) & 15 \\
1993H & 1.70(0.10)\tablenotemark{c} & \nodata & -46.790(6.947) & -21.166(3.053) & -27.335(0.697) & 35.324(0.093) & 51.183(1.020) & 16 \\
1994D & 1.558(0.013) & CN & -53.234(0.212) & -14.450(0.182) & -48.604(0.216) & 37.600(0.114) & 35.566(0.207) & 17,18 \\
1996X & 1.299(0.009) & CN & -53.837(0.216) & -9.917(0.400) & -51.935(0.309) & 41.193(0.271) & 43.070(0.317) & 19,20  \\
1997bp & 1.231(0.013) & \nodata & -64.847(1.884) & -3.015(1.935) & \nodata\tablenotemark{k} & 42.951(0.729) & \nodata\tablenotemark{k} & 21 \\
1997br & 1.141(0.021) & SS & -43.186(4.832) & -21.652(3.106) & \nodata\tablenotemark{k} & 40.701(0.171) & \nodata\tablenotemark{k} & 22 \\
1997do & 1.099(0.237) & BL & -57.978(0.878) & -11.934(0.186) & \nodata\tablenotemark{k} & 43.880(0.995) & \nodata\tablenotemark{k} & 23 \\
1997dt & 1.04(0.15)\tablenotemark{d} & CN & -56.255(0.148) & -8.051(0.475) & -50.003(0.632) & 41.675(0.655) & 47.482(0.538) & 23 \\
1998V & 1.150(0.025) & CN & -52.756(0.285) & -10.737(0.517) & -47.976(0.406) & 42.455(0.300) & 35.278(0.390) & 23 \\
1998aq & 1.185(0.008) & CN & -53.695(0.218) & -9.824(0.404) & -51.130(0.322) & 39.259(0.262) & 31.908(0.403) & 23,24 \\
1998bp & 1.903(0.013) & CL & -45.839(0.215) & -28.182(0.399) & -33.504(0.312) & 27.849(0.458) & 48.179(0.449) & 23 \\
1998bu & 1.014(0.008) & CN & -53.743(0.137) & -10.060(0.255) & -48.582(1.067) & 40.477(0.177) & 32.800(1.149) & 23,25,26,27 \\
1998de & 1.881(0.066) & CL & -48.672(0.294) & -28.202(0.533) & -21.675(0.329) & 38.392(0.564) & 56.127(0.232) & 23 \\
1998dh & 1.258(0.038) & BL & -55.895(0.209) & -10.240(0.388) & -47.762(0.285) & 43.403(0.257) & 45.766(0.205) & 23 \\
1998dm & 0.983(0.339) & \nodata & -53.027(0.094) & -12.198(3.126) & \nodata\tablenotemark{k} & 47.453(0.919) & \nodata\tablenotemark{k} & 23 \\ 
1998ec & 1.074(0.028) & BL & -62.305(0.121) & -4.422(0.232) & -47.332(0.591) & 43.144(0.037) & 42.297(0.887) & 23 \\
1998eg & 1.15(0.09)\tablenotemark{e} & CN & -53.627(0.317) & -11.681(0.570) & -52.964(0.531) & 37.920(0.349) & 39.311(0.394) & 23 \\
1998es & 0.745(0.013) & SS & -52.360(0.668) & -4.399(0.549) & -45.227(0.306) & 42.884(0.086) & 31.574(0.194) & 23 \\
1999aa & 0.811(0.014) & SS & -51.631(0.885) & -5.908(0.408) & -44.397(0.210) & 43.020(0.087) & 30.685(0.214) & 23,28 \\ 
1999ac & 1.241(0.036) & SS & -56.123(0.280) & -9.354(0.509) & -44.465(0.327) & 42.953(0.272) & 44.159(0.183) & 23,29 \\
1999aw & 0.814(0.018) & \nodata & -52.965(4.726) & -1.590(0.431) & \nodata\tablenotemark{l} & \nodata\tablenotemark{l} & \nodata\tablenotemark{l} & 30 \\
1999by & 1.796(0.008) & CL & -46.242(0.484) & -27.599(0.338) & -23.743(0.369) & 34.528(0.282) & 56.148(0.363) & 23 \\
1999cc & 1.567(0.102) & BL & -54.551(0.368) & -17.619(0.654) & -46.891(0.486) & 40.408(0.439) & 41.747(0.363) & 23 \\
1999cl & 1.243(0.043) & BL & -59.260(0.397) & -6.847(0.703) & -38.325(0.538) & 43.409(0.161) & 46.195(0.193) & 23 \\
1999dq & 0.973(0.030) & SS & -50.336(0.623) & -6.492(0.359) & -43.249(0.249) & 43.592(0.086) & 33.746(0.162) & 23 \\
1999ee & 0.944(0.006) & SS & -52.461(0.418) & -9.313(0.737) & -48.746(0.446) & 43.475(0.215) & 39.278(0.177) & 31 \\
1999ej & 1.446(0.018) & BL & -51.523(0.241) & -21.127(0.443) & -43.869(0.384) & 36.897(0.338) & 41.707(0.383) & 23 \\
1999gh & 1.721(0.008) & BL & -53.856(2.065) & -21.929(0.901) & \nodata\tablenotemark{k} & 39.822(3.692) & \nodata\tablenotemark{k} & 23 \\
1999gp & 1.029(0.186) & SS & -54.055(0.844) & -3.634(1.409) & -46.510(0.863) & 41.516(0.449) & 37.941(0.408) & 23 \\
2000E & 1.079(0.021) & SS & -51.938(0.250) & -7.507(0.493) & -46.913(0.479) & 42.393(0.248) & 33.210(0.667) & 32 \\
2000cf & 1.364(0.043) & \nodata & -51.376(1.230) & -13.827(0.031) & -48.438(0.101) & 41.395(0.424) & 47.646(0.232) & 23 \\
2000cn & 1.675(0.027) & CL & -51.405(0.250) & -23.634(0.078) & \nodata\tablenotemark{k} & 38.841(1.773) & \nodata\tablenotemark{k} & 23 \\
2000cx & 0.971(0.006) & SS & -52.441(0.429) & -4.527(0.755) & -41.327(0.525) & 44.369(0.285) & 29.687(0.418) & 23,33 \\
2000dk & 1.457(0.033) & CL & -50.096(0.017) & -23.257(0.389) & -39.544(0.354) & 34.195(0.406) & 38.875(0.413) & 23 \\
2000fa & 1.140(0.027) & CN & -54.210(0.551) & -9.293(1.827) & -44.606(1.594) & 40.845(1.452) & 33.094(1.348) & 23 \\
2001V & 0.743(0.034) & SS & -52.006(1.813) & -5.376(2.106) & -35.152(3.846) & 39.213(4.099) & 29.718(0.170) & 23 \\
2001ay & 0.543(0.006) & \nodata & -63.993(0.203) & -2.843(0.378) & -30.552(0.265) & 39.397(0.309) & 59.731(0.666) & 34 \\
2001el & 1.166(0.004) & CN & -52.317(0.744) & -12.737(0.230) & -50.087(0.342) & 40.832(0.272) & \nodata\tablenotemark{m} & 35 \\
2002bo & 1.260(0.007) & BL & -61.527(0.322) & -5.877(0.821) & -41.535(0.730) & 45.575(0.464) & 41.272(0.458) & 36 \\
2002cx & 1.145(0.016) & SS & -32.952(3.231) & -19.216(6.785) & -28.614(6.820) & 20.330(2.847) & 27.516(2.514) & 37 \\
2002dj & 1.08(0.05)\tablenotemark{f} & \nodata & -63.346(0.267) & -2.519(2.441) & -46.720(0.321) & 46.987(1.460) & 38.889(0.614) & 38 \\
2002el & 1.423(0.018) & \nodata & -54.749(0.349) & -15.368(0.904) & -45.108(0.855) & 37.251(0.296) & 37.339(0.567) & 39 \\
2002er & 1.301(0.009) & BL & -56.853(0.227) & -10.436(0.420) & -49.949(0.369) & 41.584(0.248) & 45.197(0.332) & 40 \\
2003cg & 1.25(0.05) & \nodata & -52.901(0.295) & -8.980(0.506) & -43.235(0.634) & 38.695(0.223) & 36.871(0.400) & 41 \\
2003du & 1.151(0.037) & CN & -54.388(0.192) & -7.664(0.360) & -53.238(0.300) & 40.229(0.287) & 39.471(0.283) & 42,43 \\
2004S & 1.210(0.016) & CN & -48.235(3.103) & -18.474(5.782) & \nodata\tablenotemark{k} & 40.250(2.452) & \nodata\tablenotemark{k} & 45 \\
2004dt & 1.299(0.002) & \nodata & -64.706(0.662) & -3.370(0.231) & -41.069(0.266) & 34.912(0.214) & 45.159(0.400) & 46 \\
2004eo & 1.417(0.004) & CL & -49.673(1.001) & -20.883(0.351) & -40.903(0.220) & 41.306(0.303) & 49.330(4.384) & 47 \\
2005bl & 1.93(0.10) & \nodata & -45.532(3.327) & -25.784(3.759) & -17.737(3.337) & 27.149(0.361) & 40.669(1.949) & 48 \\
2005cf & 1.161(0.006) & CN & -52.758(0.249) & -11.505(0.457) & -52.686(0.316) & 40.237(0.356) & 43.805(0.364) & 49 \\
2005cg & 0.942(0.048)\tablenotemark{g} & \nodata & -55.780(0.292) & -6.463(0.528) & \nodata\tablenotemark{n} & \nodata\tablenotemark{n} & \nodata\tablenotemark{n} & 50 \\
2005df & 1.116(0.013) & \nodata & -53.024(0.064) & -9.486(3.441) & -51.371(0.514) & 46.312(2.108) & 37.292(2.676) & 51 \\
2005hj & 0.743(0.165)\tablenotemark{h} & \nodata & -51.129(0.542) & -4.260(0.935) & -41.942(0.423) & 49.066(0.336) & 38.596(0.529) & 52 \\
2005hk & 1.56(0.09)\tablenotemark{e} & SS & -42.299(6.056) & -27.804(1.861) & -16.555(1.580) & 31.749(0.302) & 30.752(14.932) & 53  \\
2006gz & 0.69(0.04)\tablenotemark{i} & SS & -52.610(0.032) & -7.209(0.217) & \nodata\tablenotemark{k} & 33.375(0.492) & \nodata\tablenotemark{k} & 54  \\
2006X & 1.17(0.04)\tablenotemark{j} & BL & -66.111(0.252) & -0.616(0.195) & -25.726(0.209) & 42.110(0.088) & 38.500(0.186) & 55 \\


\enddata
\tablenotetext{a}{$\Delta m_{15}$ values were calculated by the super-stretch method from \citet{2006ApJ...641...50W} unless otherwise noted.}
\tablenotetext{b}{Designations from \citet{2009PASP..121..238B}} 
\tablenotetext{c}{$\Delta m_{15}$ from \citet{2006MNRAS.370..299H}} 
\tablenotetext{d}{$\Delta m_{15}$ from \citet{1999ApJS..125...73J}} 
\tablenotetext{e}{$\Delta m_{15}$ from \citet{2007PASP..119..360P}} 
\tablenotetext{f}{$\Delta m_{15}$ from \citet{2008MNRAS.388..971P}} 
\tablenotetext{g}{$\Delta m_{15}$ converted from stretch value, s, from \citet{2006ApJ...636..400Q} using the equation from \citet{1997ApJ...483..565P}} 
\tablenotetext{h}{$\Delta m_{15}$ converted from stretch value, s, from \citet{2007ApJ...666.1083Q} using the equation from \citet{1997ApJ...483..565P}} 
\tablenotetext{i}{$\Delta m_{15}$ from \citet{2007ApJ...669L..17H}}  
\tablenotetext{j}{$\Delta m_{15}$ from \citet{2008ApJ...675..626W}} 
\tablenotetext{k}{The 5485 \AA\ and 4570 \AA\ features show much more variance in their evolution, therefore the epoch range over which these features were fit was smaller. These SNe are missing $X_{5485}$ and $X_{4570}$ values because they did not have enough spectra within the smaller epoch range.}
\tablenotetext{l}{Due to noise or miscalibration of the spectra at +3 days, there is not enough data to fit $X_{5485}$, $X_{5150}$, and $X_{4570}$}
\tablenotetext{m}{The spectra for SN~2001el did not cover the wavelength region for this feature.}
\tablenotetext{n}{Not enough of the spectra for SN~2005bl covered the wavelength regions for $X_{5485}$, $X_{5150}$, and $X_{4570}$ for a good fit to be made.}

\tablerefs{(1) \citet{1983ApJ...270..123B}; (2) \citet{1983PASP...95..607H}; (3) \citet{1989A&A...220...83B}; (4) \citet{1987PASP...99..592P}; (5) \citet{1990A&A...237...79B}; (6) \citet{1994AJ....108.2233W}; (7) \citet{1993A&A...269..423M}; (8) \citet{1991ApJ...371L..23L}; (9) \citet{1992ApJ...384L..15F}; (10) \citet{1992AJ....103.1632P}; (11) \citet{1992ApJ...387L..33R}; (12) \citet{1993AJ....105..301L}; (13) \citet{1992AJ....104.1543F}; (14) \citet{1996MNRAS.283....1T}; (15) \citet{1993ApJ...415..589K}; (16) Wang unpublished; (17) \citet{1996MNRAS.281..263M}; (18) \citet{1996MNRAS.278..111P}; (19) \citet{1997ApJ...476L..27W}; (20) \citet{2001MNRAS.321..254S}; (21) \citet{1997AJ....114.2054A}; (22) \citet{1999AJ....117.2709L}; (23) \citet{2008AJ....135.1598M}; (24) \citet{2003AJ....126.1489B}; (25) \citet{1999ApJS..125...73J}; (26) \citet{2000MmSAI..71..299M}; (27) \citet{2000MNRAS.319..223H}; (28) \citet{2004AJ....128..387G}; (29) \citet{2005AJ....130.2278G}; (30) \citet{2002AJ....124.2905S}; (31) \citet{2002AJ....124..417H}; (32) \citet{2003ApJ...595..779V}; (33) \citet{2001PASP..113.1178L}; (34) \citet{2006PASP..118..560B}; (35) \citet{2003ApJ...591.1110W}; (36) \citet{2004MNRAS.348..261B}; (37) \citet{2003PASP..115..453L}; (38) \citet{2008MNRAS.388..971P}; (39) Wang unpublished; (40) \citet{2005A&A...436.1021K}; (41) \citet{2006MNRAS.369.1880E}; (42) \citet{2005A&A...429..667A}; (43) \citet{2007A&A...469..645S}; (44) \citet{2006Natur.443..308H}; (45) \citet{2007AJ....133...58K}; (46) \citet{2007A&A...475..585A}; (47) \citet{2008MNRAS.386.1897M}; (48) \citet{2008MNRAS.385...75T}; (49) \citet{2007A&A...471..527G}; (50) \citet{2006ApJ...636..400Q}; (51) Quain in progress; (52) \citet{2007ApJ...666.1083Q}; (53) \citet{2007PASP..119..360P}; (54) \citet{2007ApJ...669L..17H}; (55) \citet{2008ApJ...675..626W}}
\end{deluxetable}

\subsection{$X$ Versus the Epochs} \label{epoch}

\begin{figure}
  \epsscale{1.0}

  \plotone{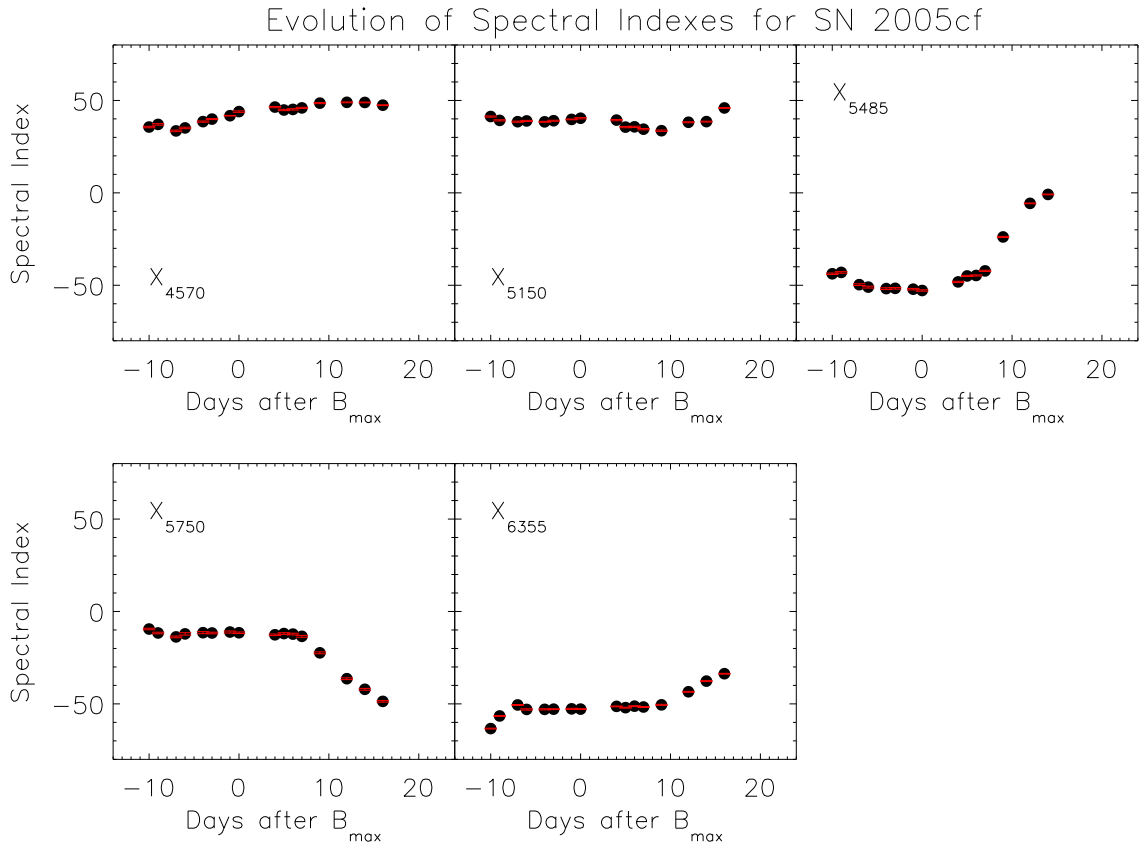}
  \caption{The temporal evolution of the $X$ indices of SN~2005cf. In
    order these are: (a) the emission feature at 4570, (b) the
    emission feature at 5150, (c) \ion{S}{2} line, (d)\ion{Si}{2} 5800
    line, (e) the \ion{Si}{2} 6150. }\label{fig:fig7}
\end{figure}

An example of the time evolution of $X$ indices is shown in Figure
\ref{fig:fig7} for SN~2005cf - a normal Type Ia supernova with $\Delta
m_{15} = 1.16$. For a spectroscopically normal supernovae like
SN~2005cf, the \ion{Si}{2} 5800 and 6150 lines exhibit little
evolution in line strength for roughly $\pm$8 days around maximum,
after which the the 5800 \AA\ line strengthens and the 6150 \AA\ line
weakens.  Similarly, around 8 days past maximum the \ion{S}{2} feature
begins to weaken until it is completely obscured by 18 days past
maximum.  The emission features at 4750 and 5150 \AA\ for this same
supernova, by contrast, shows comparatively little time evolution.

Analysis of the time evolution is complicated by occasional large gaps
between epochs and the need to occasionally track spectral features
manually due to the decreasing velocity of the expanding
photosphere. Consequently, a full analysis of the temporal evolution
of the remaining supernovae in Table \ref{tab:tab3} will be analyzed
in a separate paper.


\subsection{$X$ versus $\Delta m_{15}$}\label{delta}

Figures \ref{fig:fig8} to \ref{fig:fig12} show the correlations
between $X$ and $\Delta m_{15}$ for the five spectral features that we
have adopted for our study.  The line strengths in these figures are
those computed at maximum light.  In instances where no spectrum at
maximum light exists a simple quadratic fit was made of all spectra
within 8 days of maximum (for features that are not a smoothly varying
the fit was restricted to within 5 days of maximum).  The fits were
checked for consistency and for supernovae with only two spectra
closely sampled in time, a mean was taken to avoid aberrant behavior
in the fit.  Any supernova having only a single spectrum within the
specified time range was removed, unless that spectrum was taken at
maximum.

\subsubsection{The \ion{Si}{2} 6150 \AA\ line}

\begin{figure}
  \epsscale{1.0}

  \plotone{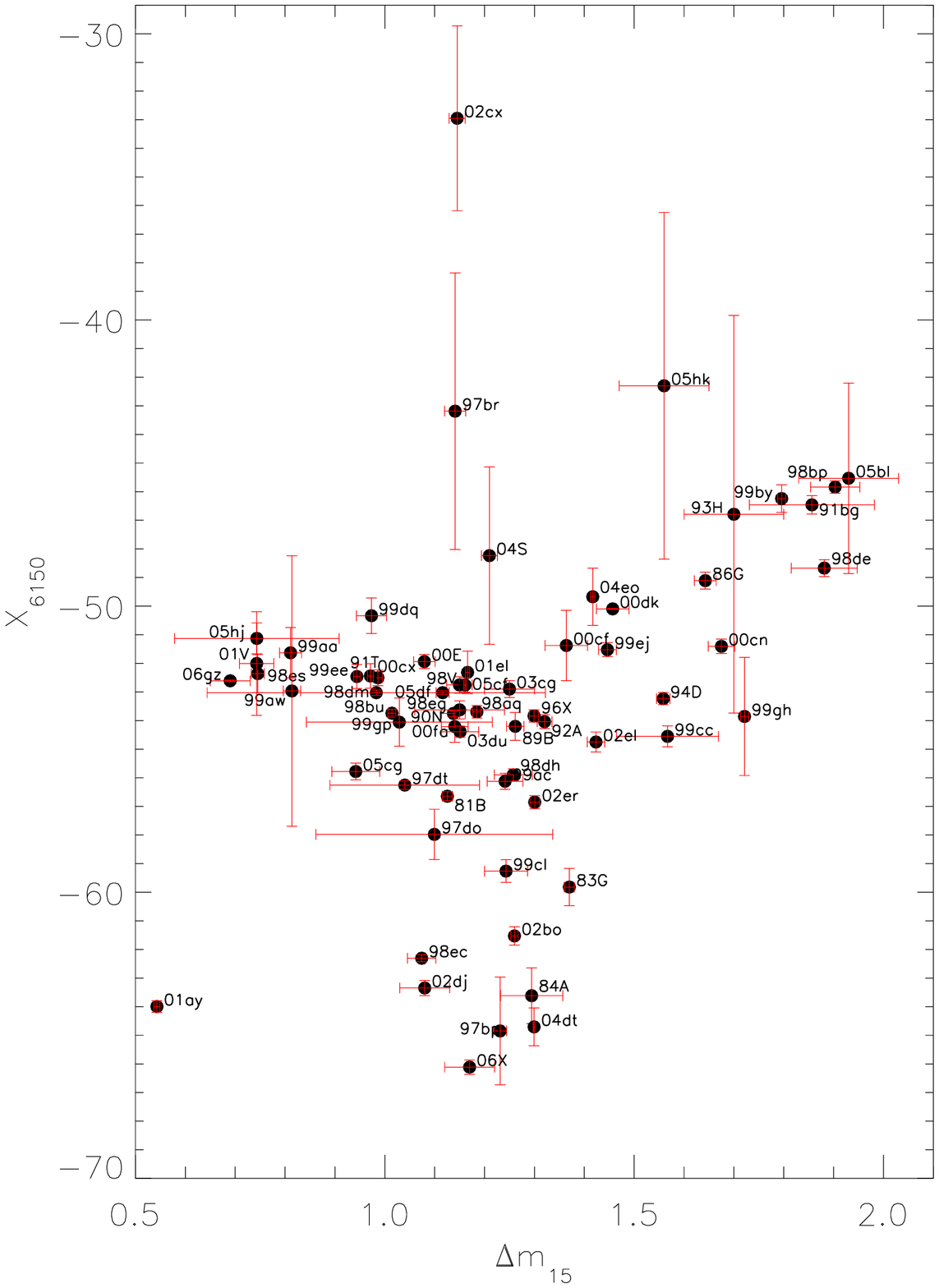}
  \caption{The correlation of the strength of \ion{Si}{2} 6150 feature
    and $\Delta m_{15}$.  }\label{fig:fig8}
\end{figure}

It has been shown previously that the strength of the \ion{Si}{2} 6150
\AA\ line is not tightly correlated with the intrinsic brightness
\citet{2006MNRAS.370..299H}.  Figure \ref{fig:fig8} confirms this
observation in the main. However, the $X$ indices do show a modest
trend of weaker spectral strength for dimmer supernovae and supernovae
with $\Delta m_{15}$ less than 1 show large variations of the
\ion{Si}{2} 6150 strength. It merits mention that the several of these
supernovae (e.g., SN~1997br, SN~2001ay, SN~2002cx, and SN~2005hk) are
deviant with respect to the majority of the sample. It has been noted
that these are all peculiar supernovae and it has been speculated that
SN~1997br, SN~2002cx, and SN~2005hk may form a group distinct from
most typical Type Ia supernovae
\citet{1999AJ....117.2709L,2003PASP..115..453L,2004PASP..116..903B,2004cetd.conf..151H,2006PASP..118..560B,2006AJ....132..189J,2007PASP..119..360P,2008ApJ...680..580S,2009PASP..121..238B}.

\subsubsection{\ion{Si}{2} 5800 \AA}

\begin{figure}
  \epsscale{1.0}

  \plotone{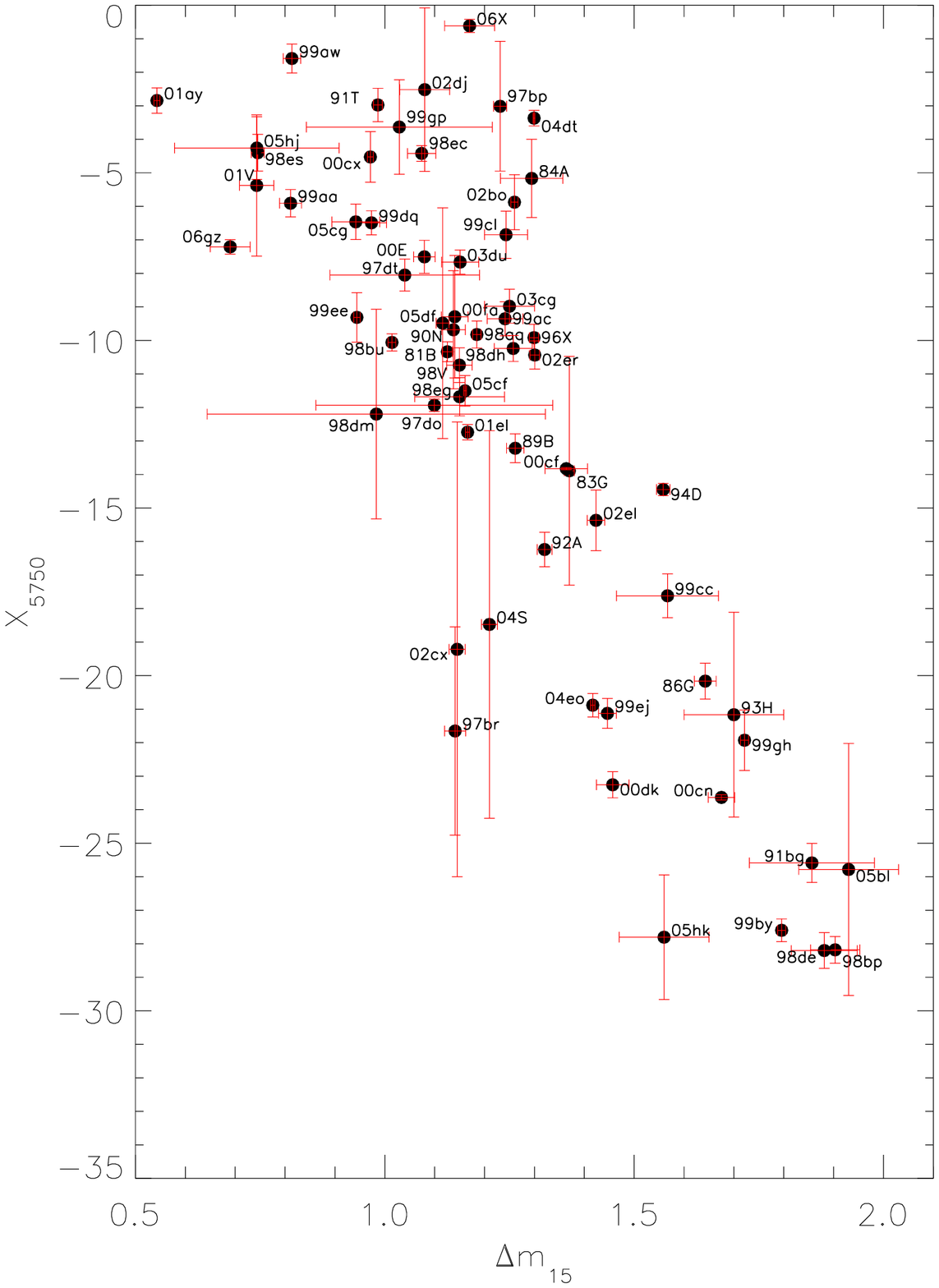}
  \caption{ The correlation of the strength of \ion{Si}{2} 5800 and
    $\Delta m_{15}$.  }\label{fig:fig9}
\end{figure}

It has previously been shown that the ratio of the strength of this
feature and that of the \ion{Si}{2} 6150 \AA\ line are well correlated
with $\Delta m_{15}$ \citet{1995ApJ...455L.147N,2006MNRAS.370..299H}.
Using this correlation, a determination of the supernovaes' maximum
luminosity may be determined on the basis of a single spectra
\citet{1998ApJ...504..935R}.  These same two features have also been
used to define Ia~SNe subgroups
\citet{2005ApJ...623.1011B,2006PASP..118..560B,2009PASP..121..238B}.

Figure \ref{fig:fig9} shows the correlations with $\Delta m_{15}$. The
$X$ index defined here for this line correlates tightly with $\Delta
m_{15}$ even without having been divided by the strength of the
\ion{Si}{2} 6150 line. Note however, that the $X$ index for this
feature is normalized by the total variance of the wavelength scales
from wavelength region between 5500 \AA\ and 6500 \AA, the variations
due to \ion{Si}{2} 6150 \AA\ line are partially included in the
derivations of the $X$ indices. The correlation between $X$ and
$\Delta m_{15}$ can be well described by a linear relation.  It may be
the case that for sub-luminous supernovae such as SN~1991bg, the
feature 5800 \AA may actually be a blend of \ion{Ti}{2} and
\ion{Si}{2} \citet{2004ApJ...613.1120G} (although this has also been
disputed \citet{2006PASP..118..560B,2008ApJ...687..456B}). Blending
would artificially enhance the line strength and may suggest more of a
correlation than in the absence of line blending. We wish to emphasize
that the $X$ index of this feature measures the total strength of this
feature and does not distinguish the physical origins of the features.

\subsubsection{\ion{S}{2} 5485}

\begin{figure}
  \epsscale{1.0}

  \plotone{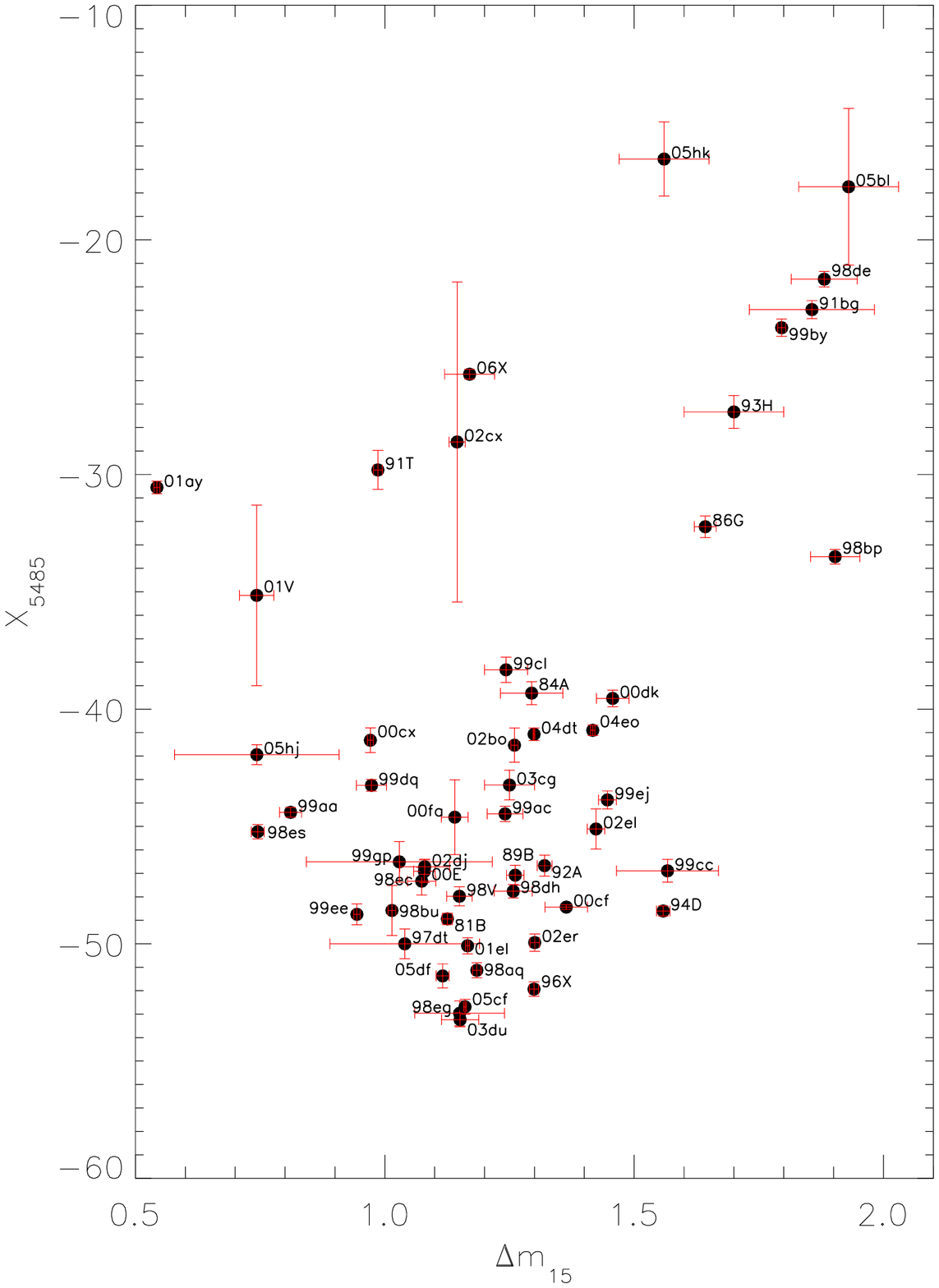}
  \caption{ The correlation of the strength of \ion{S}{2} 5485 and
    $\Delta m_{15}$.  }\label{fig:fig10}
\end{figure}

The ``w'' shaped spectral feature \ion{S}{2} 5485 is another important
line that defines an SN~Ia similar to the \ion{Si}{2} 6150 lines
\citet{2006ApJ...647..513B}. Figure \ref{fig:fig10} suggests that the
strength of this feature too may depend on $\Delta m_{15}$, though
correlation is much weaker than that for \ion{Si}{2} 5800.  The
deviant SNe are SN~1991T, SN~2001ay, SN~2005hk, SN~2006X, SN~2001V and
SN~2002cx.  SN~1997br has a similar spectral evolution to that of
SN~1991T, but it has not been shown since it has only one spectrum
within 5 days of maximum However the $X$ value ($-30$) for SN~1997br
at 4 days before maximum is consistent with the $X$ value for
SN~1991T.  Other 1991T-like SNe (SN~1998es, SN~1999aa, SN~1999dq,
SN~2000cx) have more normal values but they still are on the upper
edge of the distribution.  There is apparently some diversity within
the so-called SN~1991T-like SNe.

The \ion{S}{2} line is generally much stronger than the \ion{Si}{2}
5800 line and is thus much easier to measure.  There is also much more
evolution within this feature so Figure 10 is restricted to spectra
taken within 5 days of maximum.

\subsection{5150 Emission Feature}

\begin{figure}
  \epsscale{1.0}

  \plotone{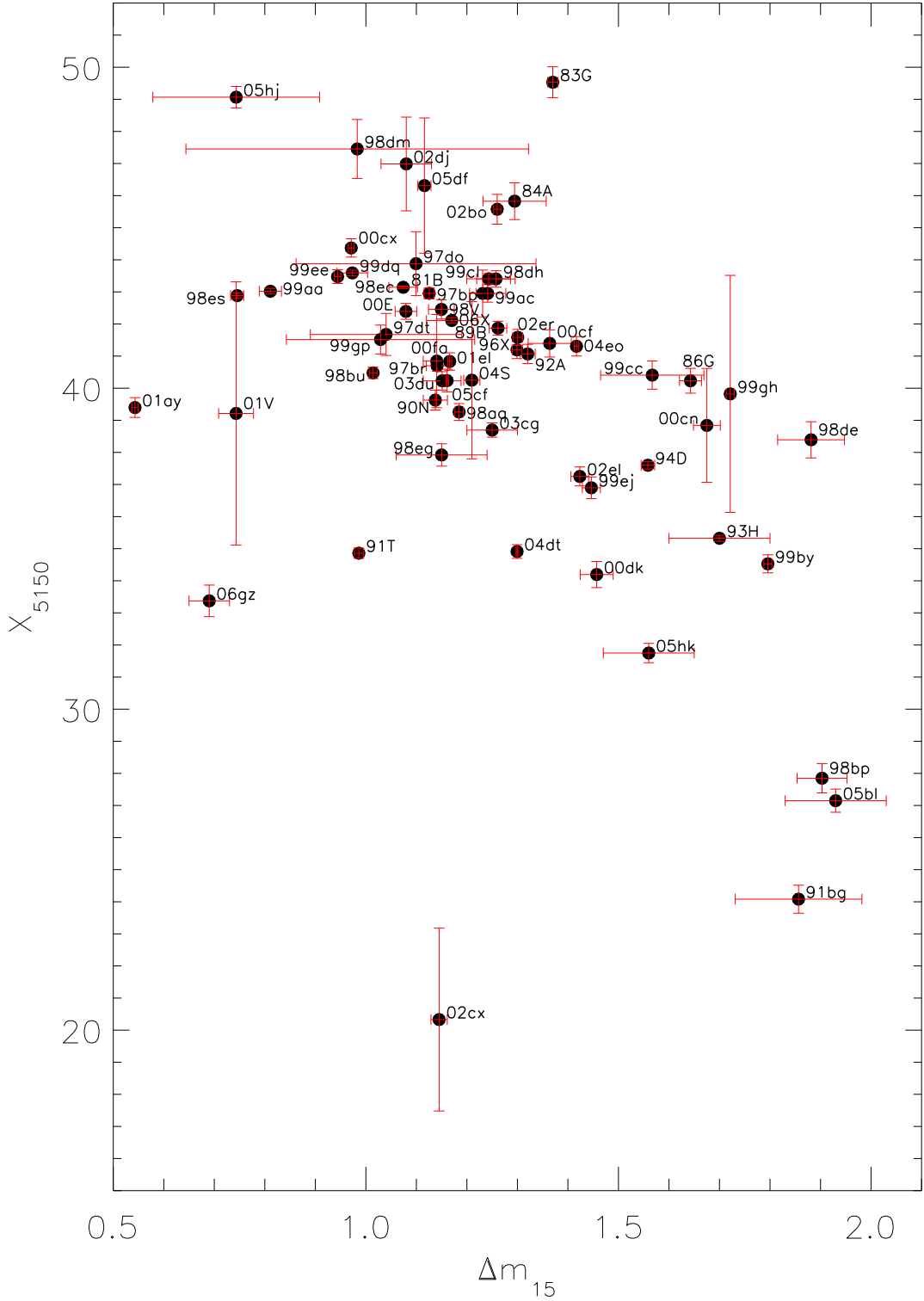}
  \caption{ The correlation of the emission peak at 5150 \AA\ and
    $\Delta m_{15}$.  }\label{fig:fig11}
\end{figure}

Figure \ref{fig:fig11} shows what might be interpreted as a slight
dependence on $\Delta m_{15}$, though the trend is not as apparent as
it is with the \ion{S}{2} line.  The $X$ values do decrease notably
with $\Delta m_{15}$ for the most sub-luminous supernovae.  A majority
of the data are clustered with a large amount of scatter.  The 5150
\AA\ emission feature may not be a good indicator of decline rate.

\subsection{4570 Emission Feature}

\begin{figure}
  \epsscale{1.0}

  \plotone{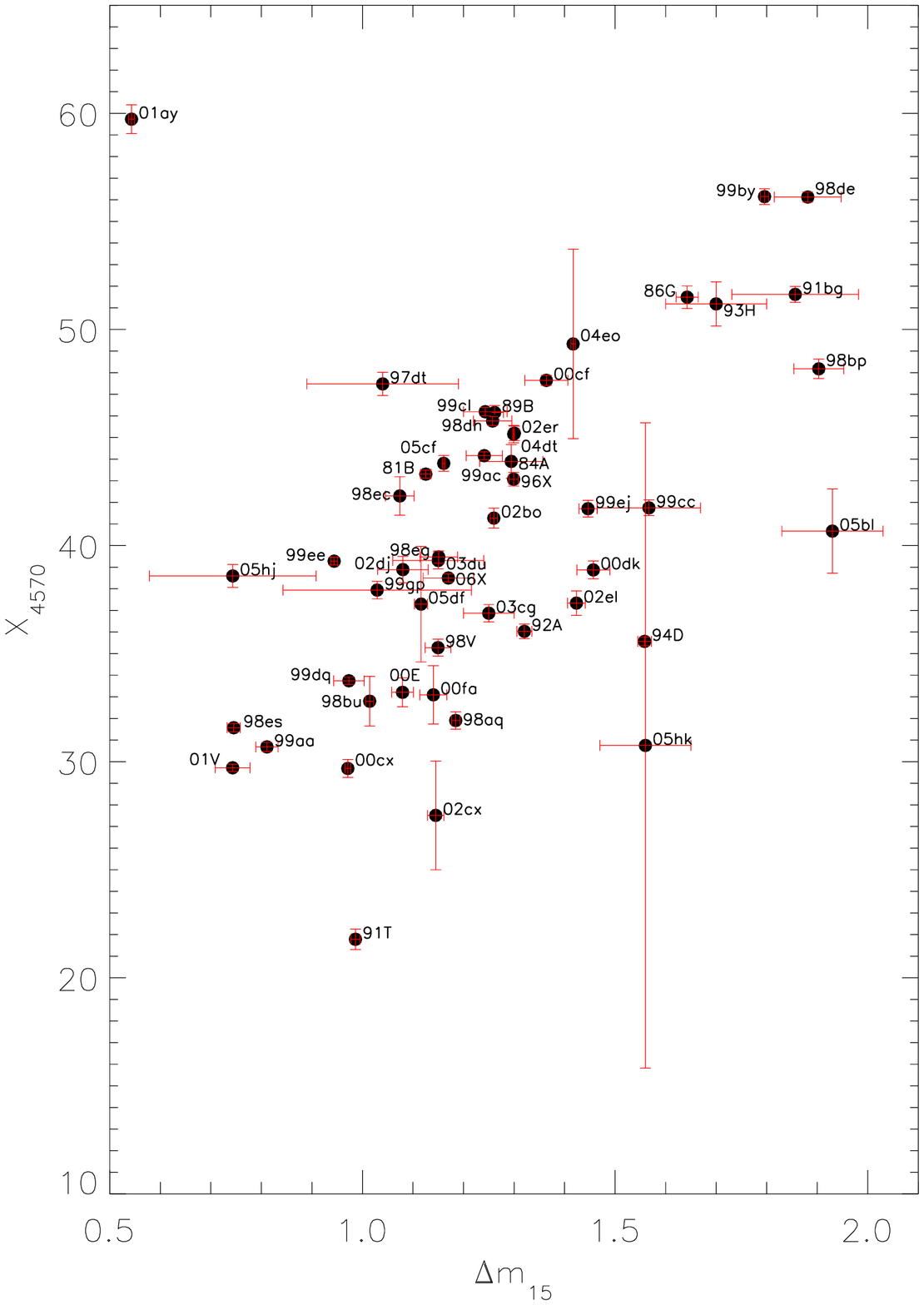}
  \caption{ The correlation of the strength of the emission peak at
    4570 \AA\ and $\Delta m_{15}$.  }\label{fig:fig12}
\end{figure}

Similar to Figure \ref{fig:fig11}, Figure \ref{fig:fig12} shows what
appears to be weak dependence on $\Delta m_{15}$ with the 4570 \AA\
emission peak. As with Figures \ref{fig:fig9} and \ref{fig:fig10} the
more deviant SNe appear on the outer edges of the distribution.  This
tendency is somewhat stronger than that found for the 5150 \AA\
feature and is in the opposite direction: the $X$ value for 4570 \AA\
is increasing with increasing $\Delta m_{15}$.  

There is wide variation in early time evolution to this feature, so
the data comprising Figure \ref{fig:fig12} is restricted to spectra
taken within 5 days of maximum. Similar to the 5150 \AA\ feature, the
utility of the 4750 \AA\ feature in specifying decline rate is
uncertain.

\subsection{Ratio Between the Emission Features at 4570 \AA\ and 5150 \AA\ }

\begin{figure}
  \epsscale{1.0}

  \plotone{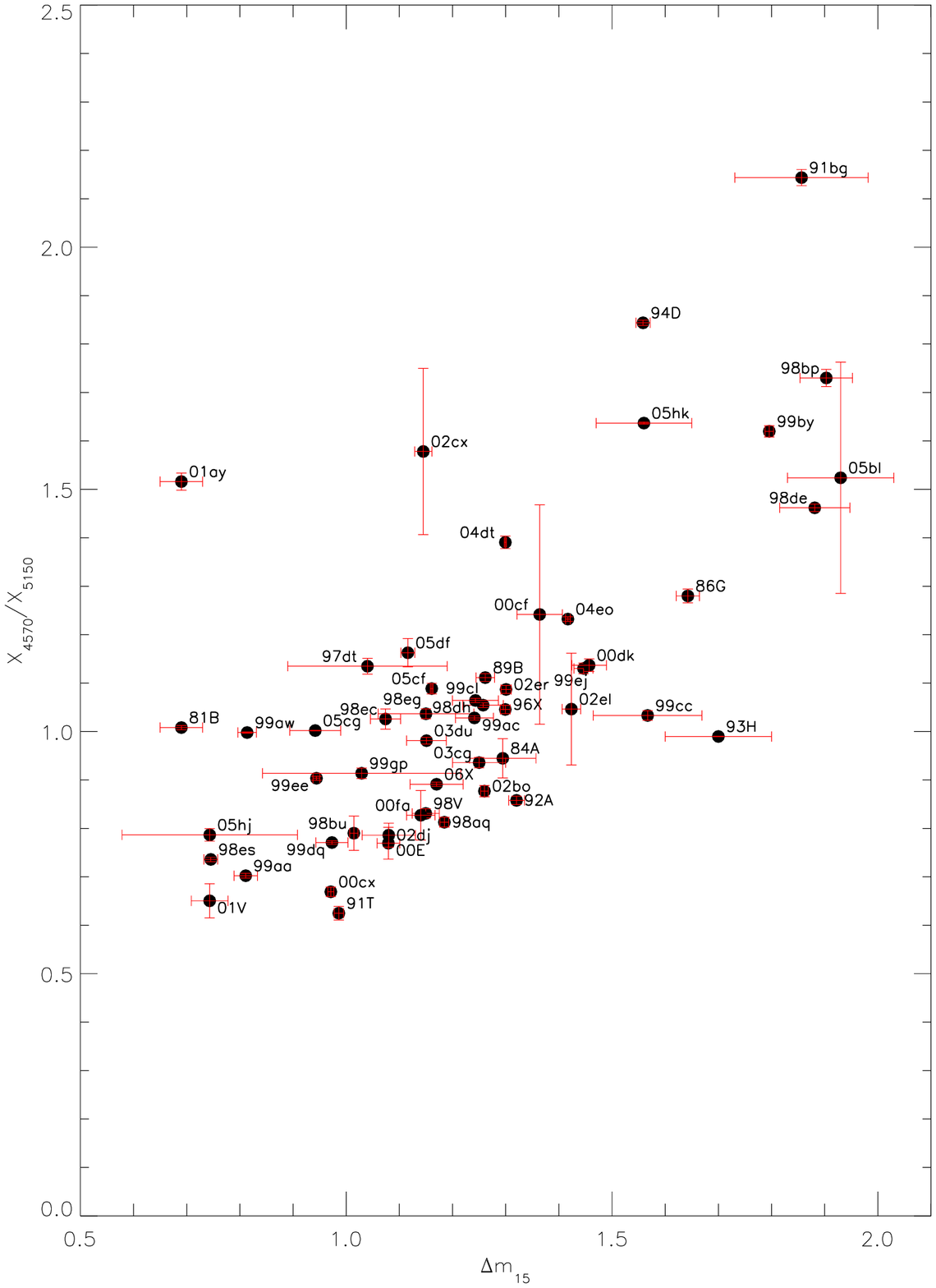}
  \caption{ The correlation of the ratio of the strength of the
    emission peak at 4570 \AA\ and 5150 \AA\ and $\Delta m_{15}$.
  }\label{fig:fig13}

\end{figure}

The suggestion of a correlation between $\Delta m_{15}$ and the ratio
of these two features appears in Figure \ref{fig:fig13}, though it is
also weak, particularly for the fast decliners.  It appears that the
4570 \AA\ feature has a stronger effect on this ratio than the 5150
\AA\ feature. 

Due to the restriction on the 4570 \AA\ feature, the same restriction
to spectra within 5 days of maximum is applied to Figure
\ref{fig:fig13}.  This ratio may be a useful parameter for the slower
decliners but not for fast decliners.


\subsection{Extinction}

To explore the impact of reddening on the spectral indices, each
spectra in the supernova sample in Table \ref{tab:tab3} was reddened.
Four values for E(B-V) were chosen: -0.25, 0.25, 0.5, and 1.0. An
identical procedure to that described in Section \ref{sec:ba} was
applied to these reddened spectra and spectral indexes were
recalculated. 

\begin{figure}
  \epsscale{1.0}

  \plotone{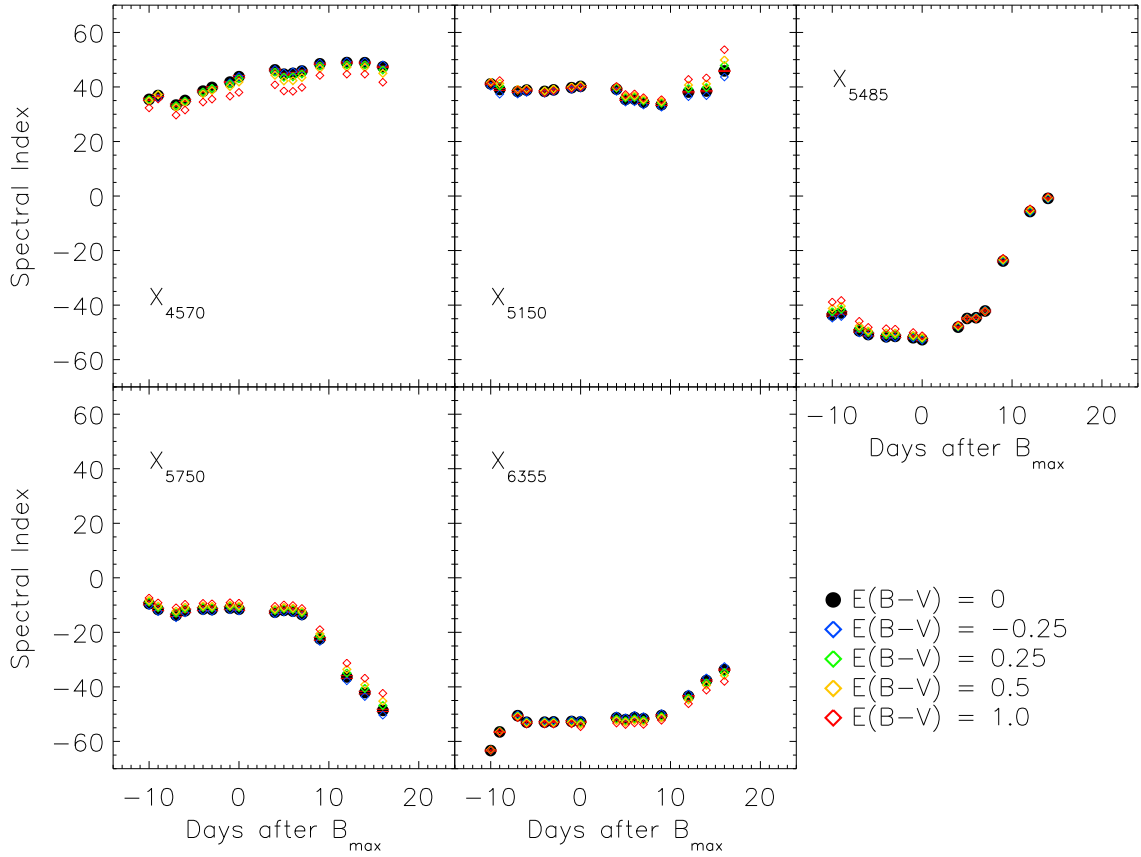}
  \caption{Same as Figure \ref{fig:fig7}, but with the spectral
    indices recomputed separately for each of the 4 E(B-V)
    values.}\label{fig:fig14}

\end{figure}

\begin{figure}
  \epsscale{1.0}

  \plotone{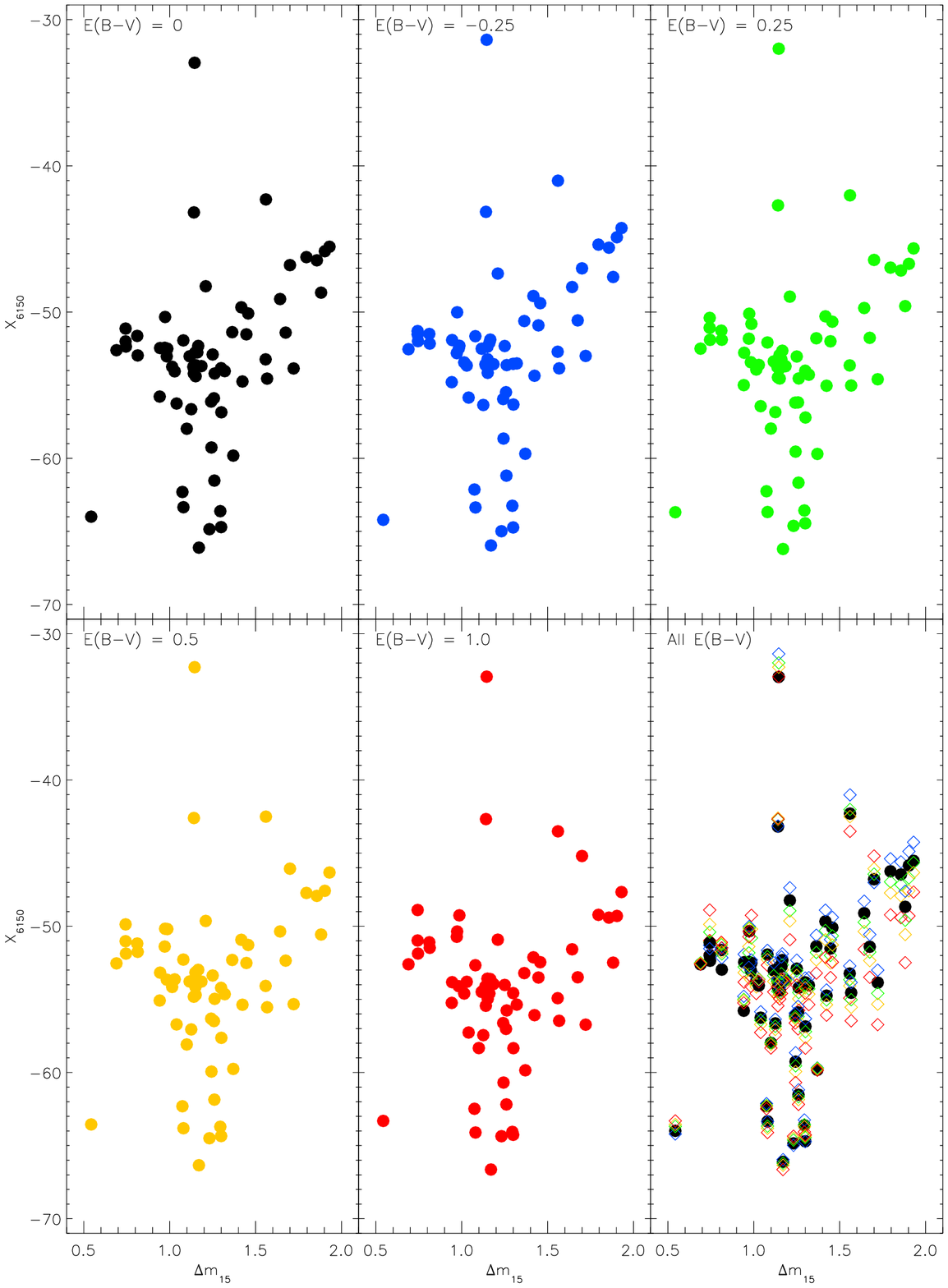}
  \caption{Same as Figure \ref{fig:fig8}, but with the addition of
    four panels each displaying the spectral indices for four values
    of E(B-V). The lower right panel overlays data from the first
    five panels. }\label{fig:fig15}

\end{figure}

Figures \ref{fig:fig14} and \ref{fig:fig15} are similar to Figures
\ref{fig:fig7} and \ref{fig:fig8}, but now include the effects of
various values of E(B-V).  Figures \ref{fig:fig14} and \ref{fig:fig15}
together demonstrate that reddening has only a minor effect on
spectral indices and is, in any event, an effect that can be
corrected.

\section{Conclusions}

Efforts to correlate the intrinsic brightness of Type Ia supernovae
with spectral line ratios or other methods related to feature strength
are considerably complicated by line blending and noise. We introduce
a comparatively new technique to astronomical image processing based
on an \`a trous wavelet decomposition and use it to extract spectral
strengths of Type Ia supernovae. In a straightforward manner repeated
application of the \`a trous generates successively smoother
scales. The lowest scale can be identified with noise; a smooth
residual results from truncating the algorithm at the highest
scale. The intermediate scales are those that can be identified with
spectral features and combining several of these intermediate scales
provides a robust measure of spectral strength without having to
wrestle with integration limits or a definition of the
continuum. Monte Carlo methods in conjunction with a very high SNR
Type Ia supernova spectrum allows for correction of the spectral
indices of supernovae with lower SNR, even those whose SNR approach
one. These same methods readily permit error bars to be assigned to
the spectral indices. The result is a definition of spectral line
strength that is applied in this paper to the temporal evolution of
spectral line strengths and the correlation of important spectral
features with $\Delta m_{15}$. These indices are also shown to be
largely impervious to reddening. A more robust spectral line strength
like that developed here is likely to advance the identification or
study of Type Ia subtypes or permit the construction of Type Ia
templates with stretches that differ from unity. The latter effort, in
particular, is directly related to our work on photometric redshift
estimation for the next generation of ground- and spaced-based survey
telescopes.

\end{document}